\documentclass[prd,showpacs,nofootinbib,twocolumn,amsmath,amssymb]{revtex4}
\newcommand{\figwidth}{0.95\linewidth}
\pdfoutput=1  
\newcommand{\defaultcolumngrid}{\twocolumngrid}

\usepackage[final]{graphicx}
\usepackage[T1]{fontenc}
\usepackage{lmodern}
\usepackage[
  hyperfootnotes=false,
  urlbordercolor={0.9 0.5 0.5},
  citebordercolor={0.5 0.9 0.5},
  linkbordercolor={0.7 0.7 1}]{hyperref}

\usepackage{xcolor}

\newcommand{\eprint}[1]{\striparxiv #1:}
\def\striparxiv#1:#2:{\href{http://www.arXiv.org/abs/#2}{#2}}

\newcommand{\bfA}{\mathbf{a}}
\newcommand{\bfB}{\mathbf{b}}
\newcommand{\bfC}{\mathbf{c}}
\newcommand{\bfy}{\mathbf{y}}

\newcommand{\cO}{\mathcal{O}}
\newcommand{\cN}{\mathcal{N}}
\newcommand{\cY}{\mathcal{Y}}
\newcommand{\cF}{\mathcal{F}}

\newcommand{\NB}{\nobreakdash}
\newcommand{\p}{\partial}

\newcommand{\OO}[1]{\mathcal{O}\bigl(#1\bigr)}
\newcommand{\vev}[1]{\langle#1\rangle}
\newcommand{\Lag}{\mathcal{L}}

\DeclareMathOperator{\F}{{}_2F_1}
\DeclareMathOperator{\diag}{diag}
\newcommand{\BesselJ}[1]{\,\mathrm{J}_{#1}}
\newcommand{\BesselY}[1]{\,\mathrm{Y}_{#1}}
\newcommand{\tint}{{\textstyle\int}}

\providecommand{\acro}[1]{\small{#1}\@} 
\newcommand{\Poincare}{Poincar\'e}
\newcommand{\AdSCFT}{\acro{AdS}/\acro{CFT}}

\begin{document}
\title{Flavors in an expanding plasma}
\author{Johannes Gro\ss{}e}
\author{Romuald A.\ Janik}
\author{Piotr Sur\'owka}
\affiliation{Institute of Physics, Jagiellonian University, Reymonta 4, 30-059 Krak\'{o}w, Poland}
\date{October\ 17, 2007}

\begin{abstract}
   We consider the effect of an expanding plasma on probe matter by
   determining time-dependent D7 embeddings in the holographic dual of
   an expanding viscous plasma. We calculate the chiral condensate and
   meson spectra including contributions of viscosity.  The
   chiral condensate essentially confirms the expectation from the
   static black hole. For the meson spectra we propose a scheme
   that is in agreement with the adiabatic approximation. New 
   contributions arise for the vector mesons at the order of the
   viscosity terms.
\end{abstract}
\preprint{arXiv:0709.3910}
\maketitle


\section{Introduction}
The study of the properties of strongly interacting quark-gluon plasma
(\acro{QGP}) \cite{review} is one of the most investigated research
topics in recent years. The interest is fueled on the one hand by
direct experimental questions related to the properties of \acro{QGP}
produced at \acro{RHIC} and on the other hand by the possibility of
studying analytically the non-perturbative properties of plasmas in
various, mostly supersymmetric, gauge theories using the \AdSCFT\
correspondence \cite{adscft}.  Although there does not exist so far a
direct counterpart of the \acro{QCD} plasma, the study of exact
properties of similar systems not based on \emph{ad-hoc}
phenomenological models may increase our understanding of the
properties of the \acro{QCD} plasma.  This hope has been
substantiated, e.g., by the discovery of a universal strong coupling
shear viscosity to entropy ratio, which is valid for a wide range of
different gauge theories \cite{son,other}.

A lot of work has been initially undertaken for a static plasma system
with a fixed constant temperature. After the earliest investigations
of transport coefficients various other observables have been
discussed related to the properties of fundamental flavors in the
thermal medium such as drag force calculations
\cite{Gubser:2006bz,Herzog:2006gh}, jet-quenching
\cite{Liu:2006ug,Bertoldi:2007sf,Cotrone:2007qa}, meson spectrum
\cite{KMMW,ErdmengerEvans}, meson melting \cite{Hoyos:2006gb},
thermodynamics of fundamental flavors \cite{Mateos:2007vn}.

On the other hand, since the experimentally produced plasma is always
a non-static expanding system, it was tempting to extend the \AdSCFT\
investigations to such a time-dependent dynamical setting.
Qualitative duals for thermalization and cooling have been suggested
in \cite{nastase,zahed}, a quantitative framework for studying
boost-invariant expansion of a plasma system in $\cN=4$ \acro{SYM} has
been proposed in \cite{JP1}. Subsequent work within this framework
include
\cite{SJSIN,Heller:2007qt,RJ,JP2,BAK,SJSIN2,Kajantie,Siopsis,Kovchegov},
which concentrated on the details of the dynamics of the expanding
plasma in \emph{pure} $\cN=4$ \acro{SYM} theory.

The main motivation for this work was to bring together the two lines
of investigation and to study an expanding plasma in the $\cN=4$
\acro{SYM} theory with fundamental flavors. The flavor system is
represented by an embedding of a D7 brane in the dual geometry of the
expanding plasma, which is -- in an essential way -- time-dependent.
Such time-dependent embeddings have not been considered so far in the
literature on flavor systems. In this paper we would like to
concentrate on the features of the system that are the direct
analogues of the corresponding properties studied in the static case.
In particular we find the (time-dependent) embedding, we find the
time-dependence of the chiral condensate including leading viscosity
effects and we describe the behavior of mesonic modes in the scalar
and vector channels. Finally we perform a holographic renormalization
of the D7 action density. We will show that the chiral condensate,
meson spectra and action density are 
compatible with the adiabatic approximation, i.e.\ their leading
contribution agrees with the result obtained from the static AdS/Blackhole
by na\"ively assuming Bjorken scaling.

As a word of caution let us emphasize that the gauge theory that we
are considering does not exhibit spontaneous chiral symmetry breaking
so from this point of view differs substantially from
\acro{QCD}. However, since this is the first investigation of a flavor
system in an expanding plasma, we chose to deal with the simplest
theoretical setting, which is the only system for which the dual
expanding geometry is known so far. Ultimately one would like to
extend these investigations to theories that exhibit chiral symmetry
breaking.

The plan of this paper is as follows. We will first review some basic
facts about viscous hydrodynamics from field theory perspective and
its implementation in \AdSCFT. In the next section we will
perturbatively determine the time-dependent embedding of a D7 brane in
that geometry and determine the chiral quark condensate. Thereafter,
we determine meson spectra from fluctuations about the embedding.
Moreover, we give the regularized D7 action and show that it is again
compatible with expectations from adiabatic considerations.

\subsection{Boost invariant kinematics}
An interesting kinematical regime of the expanding plasma is the
so-called central rapidity region. There, as was suggested by Bjorken
\cite{Bjorken}, one assumes that the system is invariant under
longitudinal boosts. This assumption is in fact commonly used in
realistic hydrodynamic simulations of \acro{QGP} \cite{hydro}. If in
addition we assume no dependence on transverse coordinates (a limit of
infinitely large nuclei) the dynamics simplifies enormously.

In order to study boost-invariant plasma configurations it is
convenient to pass from Minkowski coordinates $(x^0,x^1, x_\perp)$ to
proper-time/spacetime rapidity ones $(\tau,y,x_\perp)$ through
\begin{align}
x^0&=\tau \cosh y & x^1&=\tau \sinh y
\end{align}
The object of this work is to describe the spacetime dependence of the
energy-momentum tensor of a boost-invariant plasma in $\cN=4$
\acro{SYM} theory at strong coupling. The symmetries of the problem
reduce the number of independent components of $T_{\mu\nu}$ to three.
Energy-momentum conservation $\partial_\mu T^{\mu\nu}=0$ and
tracelessness $T^\mu_\mu=0$ allows to express all components in terms
of just a single function -- the energy density $\varepsilon(\tau)$ in
the local rest frame.  Explicitly we have \cite{JP1}
\begin{align}
T_{\tau\tau} &= \varepsilon(\tau) \\
T_{yy} &= -\tau^2 \left( \varepsilon(\tau)+\frac{d}{d\tau} \varepsilon(\tau) \right) \\
T_{xx} &= \varepsilon(\tau)+\frac{1}{2} \tau \frac{d}{d\tau} \varepsilon(\tau)
\end{align}
Gauge theory dynamics should now pick out a definite function
$\varepsilon(\tau)$.

\subsection{Viscous hydrodynamics}
The object of a hydrodynamic model is to determine the spacetime
dependence of the energy-momentum tensor for an expanding (plasma)
system.  The simplest dynamical assumption is that of a perfect fluid.
This amounts to assuming that the energy momentum has the form
\begin{equation}
\label{e.perfect}
T_{\mu\nu}=(\varepsilon+p)u_\mu u_\nu+p \eta_{\mu\nu}
\end{equation}
where $u^\mu$ is the local 4-velocity of the fluid ($u^2=-1$),
$\varepsilon$ is the energy density and $p$ is the pressure. In the
case of $\cN=4$ \acro{SYM} theory that we consider here $T^\mu_\mu=0$
and hence $\varepsilon=3p$.  The equation of motion that one obtains
from energy conservation in the boost-invariant setup is
\begin{equation}
\partial_\tau \varepsilon = -\frac{\varepsilon+p}{\tau} \equiv -\frac{4}{3} \frac{\varepsilon}{\tau}
\end{equation}
whose solution is the celebrated Bjorken result
\begin{equation} \label{e.bjorken}
\varepsilon=\frac{\varepsilon_0}{\tau^{\frac{4}{3}}}
\end{equation}
Once one wants to include dissipative effects coming from shear
viscosity, the description becomes more complex. In a first
approximation one adds to the perfect fluid tensor a dissipative
contribution $\eta (\nabla_\mu u_\nu+\nabla_\nu u_\mu)$, where $\eta$
is the shear viscosity of the fluid. The resulting equations of motion
get modified to
\begin{equation} \label{e.epsdeq}
\partial_\tau \varepsilon = -\frac{4}{3} \frac{\varepsilon}{\tau} +\frac{4\eta}{3\tau^2}
\end{equation}
Note that in the above equation the shear viscosity is generically
temperature dependent ($\eta \propto T^3$ in the $\cN=4$ case) and
hence $\tau$ dependent. In order to have a closed system of equations
we have to incorporate this dependence through
\begin{equation}
\label{e.etaeps}
\eta = \eta_0 \cdot \varepsilon^{\frac{3}{4}}
\end{equation}
with $\eta_0$ being some numerical coefficient.
The resulting energy density satisfying \eqref{e.epsdeq} gets modified from
\eqref{e.bjorken} to 
\begin{equation} \label{e.epstauhydro}
  \varepsilon(\tau)
    =
      \frac{\varepsilon_0}{\tau^{\frac{4}{3}}} \cdot
      \Bigl( 1 -\frac{2\eta_0}{\varepsilon_0^{\frac{1}{4}} \tau^{\frac{2}{3}}}
               + \dots \Bigr).
\end{equation}

\subsection{AdS/CFT description of an expanding boost-invariant plasma}
In \cite{JP1,RJ,SJSIN,Heller:2007qt} the program of perturbatively
determining the dual geometry to the viscous hydrodynamic model
discussed in the previous section was carried out. The non-compact
part of the metric takes the form
\begin{align} \label{e.abcmetric}
\frac{ds^2}{L^2}
  &= \frac{1}{z^2} \tilde{g}_{\mu\nu}dX^\mu\,dX^\nu  + \frac{dz^2}{z^2} \\
\tilde{g}_{\mu\nu} dX^\mu\,dX^\nu
  &= \left( -e^{\mathcal{A}(z,\tau)} d\tau^2+e^{\mathcal{B}(z,\tau)} \tau^2 dy^2 +e^{\mathcal{C}(z,\tau)} dx_\perp^2 \right) \notag
\end{align}
with $z$ the holographic direction. The energy-momentum tensor
is related to the fourth order term of $\tilde{g}_{\mu\nu}$ expanded in $z$ \cite{Skenderis}.
\begin{equation}\label{e.epsa}
\begin{aligned}
\vev{T_{\mu\nu}}
   &= \frac{N_c^2}{2\pi^2} \lim_{z\to0} \frac{1}{z^4} ( \tilde{g}_{\mu\nu} - \eta_{\mu\nu}) \\
\implies \varepsilon(\tau)
  &= - \frac{N_c^2}{2\pi}\lim_{z \to 0} \frac{\mathcal{A}(z,\tau)}{z^4}
\end{aligned}
\end{equation}
Of course it is too difficult to perform this construction for
arbitrary functions $\varepsilon(\tau)$.  What has been performed in
practice is an expansion of $\varepsilon(\tau)$ for (large)
proper-times $\tau$ \cite{JP1,SJSIN,RJ,Heller:2007qt}.

Each subsequent term in the expansion can then be determined by
requiring regularity of the square of the Riemann tensor
order-by-order.  In \cite{JP1} it was shown that in order to study the
large proper-time limit of the metric one is led to introduce a
scaling variable
\begin{equation}
v=\frac{z}{\tau^{\frac{1}{3}}}\varepsilon_0^\frac{1}{4}
\end{equation}
and take the limit $\tau \to \infty$ with $v$ fixed. For the
discussion of subleading terms in the metric an expansion around this
limit was performed,
\begin{equation}
\label{e.aexp}
\mathcal{A}(z,\tau)=a_0(v)+a_1(v) \frac{1}{\varepsilon_0^\frac{1}{4}\tau^{\frac{2}{3}}} + \dots,
\end{equation}
and similarly for the other coefficients.  The leading and first
subleading coefficients are given by

\begin{align}
a_0(v) &= \ln\tfrac{(1-v^4/3)^2}{1+v^4/3}, &
a_1(v) &= 2\eta_0 \tfrac{(9+v^4)v^4}{9-v^8}, \label{e.coeffs}\\
b_0(v) &= \ln (1+v^4/3), &
b_1(v) &= -2 \eta_0 \tfrac{v^4}{3+v^4} + 2\eta_0 \ln\tfrac{3-v^4}{3+v^4}, \nonumber \\
c_0(v) &= \ln(1+v^4/3), &
c_1(v) &= -2\eta_0 \tfrac{v^4}{3+v^4} -\eta_0 \ln \tfrac{3-v^4}{3+v^4}.\nonumber
\end{align}
$\eta_0$ is an undetermined integration constant (which has the
physical interpretation as the coefficient of shear viscosity). It can
be fixed from non-singularity of the metric; the resulting value
\begin{equation} \label{e.eta0}
  \eta_0=\frac{1}{2^{\frac{1}{2}} 3^{\frac{3}{4}}}
\end{equation}
is in accord with the known viscosity coefficient of $\cN=4$
\acro{SYM}, $\eta/s = 1/4\pi$, calculated in the static case
\cite{son}.

From the leading order terms in \eqref{e.coeffs}, using the similarity
with the static black hole metric, we may read off the temperature
from the position of the horizon. As discussed in \cite{SJSIN}, to
this order, the result
\begin{equation} \label{e.T}
T(\tau)=\left(\frac{4\varepsilon_0}{3}\right)^{\frac{1}{4}} \frac{1}{\pi  \tau^{\frac{1}{3}}} \left[ 1 - \frac{\eta_0}{2 \varepsilon_0^{1/4} \tau^{2/3}} \right]
\end{equation}
is compatible with the Stefan--Boltzmann law. This can be explicitly
checked by extracting the energy density from the metric expanded in
the scaling limit up to $\OO{\tau^{-2}}$ through (\ref{e.epsa}). We
obtain
\begin{equation} \label{e.epstau}
  \varepsilon(\tau)
    = \frac{N_c^2}{2\pi^2} \cdot
      \frac{\varepsilon_0}{\tau^{\frac{4}{3}}} \cdot
      \Bigl( 1 -\frac{2\eta_0}{\varepsilon_0^{\frac{1}{4}} \tau^{\frac{2}{3}}}
               + \dots \Bigr),
\end{equation}
which is consistent with viscous hydrodynamic evolution \eqref{e.epstauhydro}.

\section{Time-dependent D7 embedding}
\subsection{D7 embeddings}
Conventional \AdSCFT\ describes $\cN=4$ \acro{SYM} theory, which has
only fields in the adjoint representation, since all fields sit in the
same supermultiplet containing the gauge field. A standard method
\cite{KarchKatz} for the introduction of quenched matter into the
correspondence is by adding probe D7 branes. Strings stretching
between the probe and the original D3 stack generating the
$AdS_5\times S^5$ background give rise to an $\cN=2$ hypermultiplet in
the fundamental representation.

Let us consider geometries $M_5 \times S^5$ with $M_5$ asymptotically
$AdS_5$ and line element
\begin{align}
  ds_{10}^2 &= \frac{r^2}{L^2} ds_4^2 + \frac{L^2}{r^2} dr^2 + d\Omega_5^2 \notag \\
    &= \frac{r^2}{L^2} ds_4^2 + \frac{L^2}{r^2} d\rho^2 + \rho^2 d\Omega_3^2 + (dX^8)^2 + (dX^9)^2
\end{align}
with $r^2 = \sum (X^i)^2 = \rho^2 + (X^8)^2 + (X^9)^2=1/z^2$.

Placing a D7 probe parallel to the first eight coordinates, its
position and shape are described by the coordinates $X^8$ and $X^9$.
Manifestly, the D7 brane breaks the $SU(4) \simeq SO(6)$ symmetry of
the internal manifold to $SO(4)\times SO_{89}(2) \simeq SU(2)_L \times
SU(2)_R \times U(1)_R$.  The $SO(2)$ symmetry in the 8,9\NB-plane can
be used to rotate the embedding of the D7 to
\begin{align}
  X^8 &= 0,&
  X^9 &= \Phi(\rho,\tau),
\end{align}
such that one scalar field is sufficient for a complete description of
the embedding.

The embedding $\Phi(\rho,\tau)$ is determined from the D7 action,
\begin{align} \label{d7action}
   S_{D7} &= \mu_7 \int d^8\xi\, e^{-\varphi} \sqrt{ \det P[g]_{ab} + F_{ab} }  \notag \\
       & \qquad + \mu_7 \int d^8\xi\, P[(\p_\rho C_4)] \varepsilon^{\alpha\beta\gamma} A_\alpha \p_\beta A_\gamma,
\end{align}
where $P[...]$ denotes the pull-back to the world-volume.

While $AdS_5 \times S^5$ permits constant embeddings $\Phi\equiv m$,
more general geometries require a non-trivial profile $\Phi(\rho)$. To
our knowledge the present article is the first-time that also
time-dependence for the embedding has been considered in the context
of flavored \AdSCFT.

Close to the boundary the embedding behaves as
\begin{align}
  \Phi \xrightarrow[\rho\to\infty]{} m + \frac{c}{\rho^2} + ...,
\end{align}
where $m$ and $c$ are related to the bare quark mass $m_q$ and chiral
condensate $\vev{\cO}$ by \cite{Mateos:2007vn}\footnote{To be more
  precise these are the mass of the $\cN=2$ hypermultiplet and the
  vacuum expectation value of the operator $\cO = \bar\psi \psi +
  q^\dagger \Phi q + m_q q^\dagger q$.}
\begin{align}
  m_q &= \frac{m}{2\pi\alpha'},\\
  \vev{\cO} &= -\frac{N_f N_c}{(2\pi\ell_s^2)^3 \lambda}\, c,
\end{align}
where $\lambda = g_\text{YM}^2 N_c = 2 \pi g_s N_c = L^4/(2\ell_s^4)$.
For the AdS/\linebreak[0]Schwarz\-schild geometry such a \acro{VEV}
forms, but vanishes for $m_q\to0$ such that there is no
\emph{spontaneous} chiral symmetry breaking associated to its
appearance \cite{ErdmengerEvans}.

The common procedure for finding the embedding is to derive the
equations of motion from \eqref{d7action} and impose the requirement
that the embedding have an interpretation as a holographic
renormalization group flow. In particular this means that $\Phi(\rho)$
should be one-valued as a function of the holographic energy scale
$r$, which requires regularity as $\rho\to0$.  (This condition is not
sufficient however as will be discussed below.)  Because of the parity
symmetry $r \mapsto -r$ of the underlying geometry, which manifests
itself as $\rho \mapsto -\rho$ invariance of the equation of motion,
embedding solutions are either parity odd, known as `black hole'
solution since they hit the horizon, or even, i.e., $\p_\rho \Phi=0$,
known as `Minkowski-type' solutions.  It is easy to show that the
condition for Minkowski-type solutions corresponds to the absence of a
conical defect on the brane \cite{KarchOBannon}.

\subsection{AdS/Schwarzschild: Adiabatic Approximation}
\begin{figure}
\includegraphics[width=\figwidth]{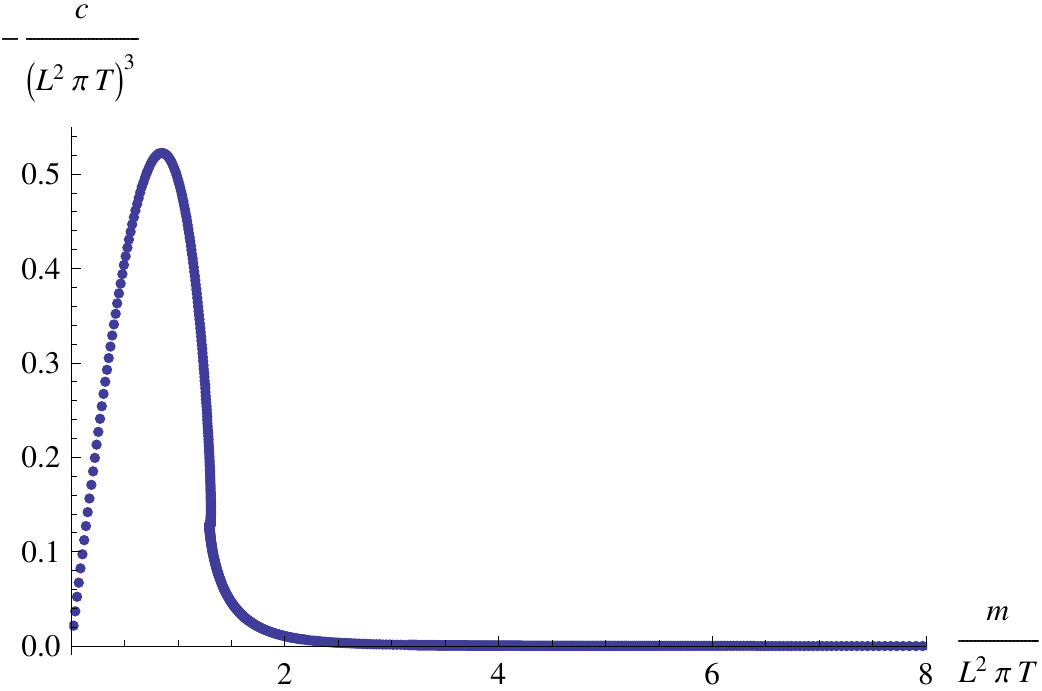}
\caption{Chiral condensate as function of the quark mass for the
  static AdS/Schwarzschild background. \label{fig:adsbh}}
\end{figure}
Figure~\ref{fig:adsbh} shows the chiral condensate as a function of
the quark mass for the AdS/Schwarzschild geometry. Since all
quantities enter the equations of motion in units of the black hole
mass $m_{bh}$ (we express everything in terms of $\pi T = 8G_N/(3\pi)
m_{bh}$), the leading order time-dependence of the chiral condensate
can be determined in an adiabatic approximation using the fact that
the perfect fluid geometry for constant $\tau$ appears like an
AdS/Schwarzschild solution with fixed temperature \eqref{e.T}.

Our numerical results provide $c/T^3 \sim (m/T)^\alpha$. $\alpha$ is
about $-6$ at $m\approx 3 (\pi T)$ and increases to $-5.2$ when going
to $m\approx 8 (\pi T)$. In other words
\begin{align}
  c \sim \tau^{-(3-\alpha)/3} \approx \tau^{-8.2/3}
\end{align}
However, as has been observed in \cite{KarchOBannon}, the numerics
determining Figure~\ref{fig:adsbh} is not particularly accurate for
large quark mass because the chiral condensate becomes small.
Therefore it is not possible to increase the quark mass till the
exponent saturates.

For large quark mass, the solutions are very far from the black hole
and become approximately constant embeddings as in the supersymmetric
scenario \cite{KarchKatz}. This suggests the following perturbative
analysis \cite{Mateos:2007vn}.  For small $\epsilon$ we seek regular
solutions
\begin{align}
  \Phi(\rho) &= m + \epsilon f(\rho)
\end{align}
of the \acro{DBI} action in AdS/Schwarzschild geometry
\cite{ErdmengerEvans}.\footnote{For conformity with our conventions,
  we have replaced their $w_6$ by $\Phi$ and the parameter $b$ by $L^2
  \pi T$, with $T$ the temperature of the black hole.}
\begin{align}
  \Lag_{D7/AdS/BH} &= \rho^3 \left( 1 - \frac{ \epsilon (L^2\pi\, T)^8 }{16
      \left(\rho ^2+\Phi(\rho)^2\right)^4}\right)
  \sqrt{1+\Phi'(\rho )}
\end{align}
We obtain
\begin{align}
  f(\rho) &= -\frac{ (L^2 \pi  T)^8 \left(3 m^4+3 \rho ^2 m^2+\rho ^4\right)}{96 m^5 \left(m^2+\rho ^2\right)^3}, \\
  \implies  \frac{c}{(L^2 \pi T)^3} &= - \frac{1}{96} \left(\frac{L^2 \pi T}{m}\right)^5. \label{adsbhvev}
\end{align}
We have written \eqref{adsbhvev} in a way that emphasizes the known
fact that the temperature can be effectively removed from the
equations by a suitable redefinition of the quark mass and condensate.
In the present context, such a rescaling is not desirable. With
\eqref{e.T} we conclude that the adiabatic estimate for the perfect 
fluid geometry is 
\begin{align}
  c &= - \frac{\varepsilon_0^2L^{16}}{54m^5} \tau^{-\frac{8}{3}}.
\end{align}
We will check in the following that this is indeed the leading
contribution for the expanding plasma geometry.

\subsection{Viscous fluid}
Restricted to the scalar $\Phi$, the action reads for our geometry
\eqref{e.abcmetric}
\begin{gather} \label{d7embedaction}
S_{D7}
  = \mathbf{N} \int d\tau\, d\rho\, \tau\,\rho^3 \,\mathbf{A} \,\sqrt{ 1 + \Phi'^2 - \mathbf{B} \frac{\dot{\Phi}^2}{(\rho^2 + \Phi^2)^2}}, \\
\begin{aligned}
  \mathbf{A} := \bigl(1-\frac{v^8}{9}\bigr) &\exp \left[ -2\eta_0 \,\varepsilon_0^{-\frac{1}{4}} \frac{v^8}{9-v^8} \tau^{-\frac{2}{3}} \right], \\
  \mathbf{B} := \frac{1+\frac{v^4}{3}}{(1-\frac{v^4}{3})^{2}} &\exp \left[ 2\eta_0\, \varepsilon_0^{-\frac{1}{4}} v^4 \frac{9+v^4}{9 - v^8} \tau^{-\frac{2}{3}} \right],
\end{aligned} \\
  v := \frac{\varepsilon_0^\frac{1}{4} L^2}{\tau^\frac{1}{3} \sqrt{\rho^2+\Phi(\rho,\tau)^2}},\\
  \mathbf{N} := N_f T_{D7} \Omega_3 V_x = \frac{1}{2} \frac{N_c N_f}{(2\pi\ell_s^2)^4 \lambda} V_x, \label{norm}
\end{gather}
where $T_{D7}=2\pi/(g_s (2\pi\ell_s)^8)$ is the D7 tension.
$\Omega_3=2\pi^2$ and $V_x=\int dy\,d^2x_\perp$ are the volume of the
unit three-sphere and spatial part of the boundary, respectively. The
latter is of course infinite.

Since the viscous fluid geometry discussed in the previous section
behaves similar to AdS/\linebreak[0]Schwarz\-schild with
time-dependent temperature, we do not expect spontaneous symmetry
breaking either, though going to small quark masses leaves the domain
of validity of the geometry and it is thus hard to make a definite
statement.  We will therefore only consider `Minkowski-type'
embeddings that avoid the horizon at the center of the geometry.

For a given quark mass, in general regularity is only possible for a
discrete set of values for the chiral condensate. In the regime under
consideration, since no phase transition occurs, we expect $c=c(m)$ to
be a one-valued function.

\begin{figure}
\includegraphics[width=\figwidth]{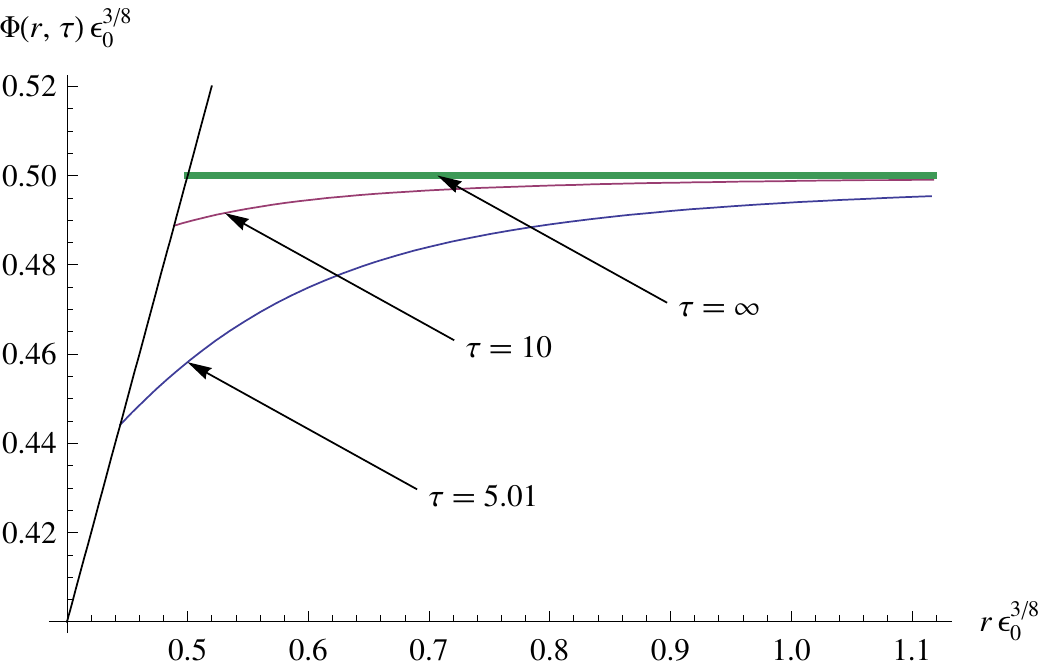}\\[1.5ex]
\includegraphics[width=\figwidth]{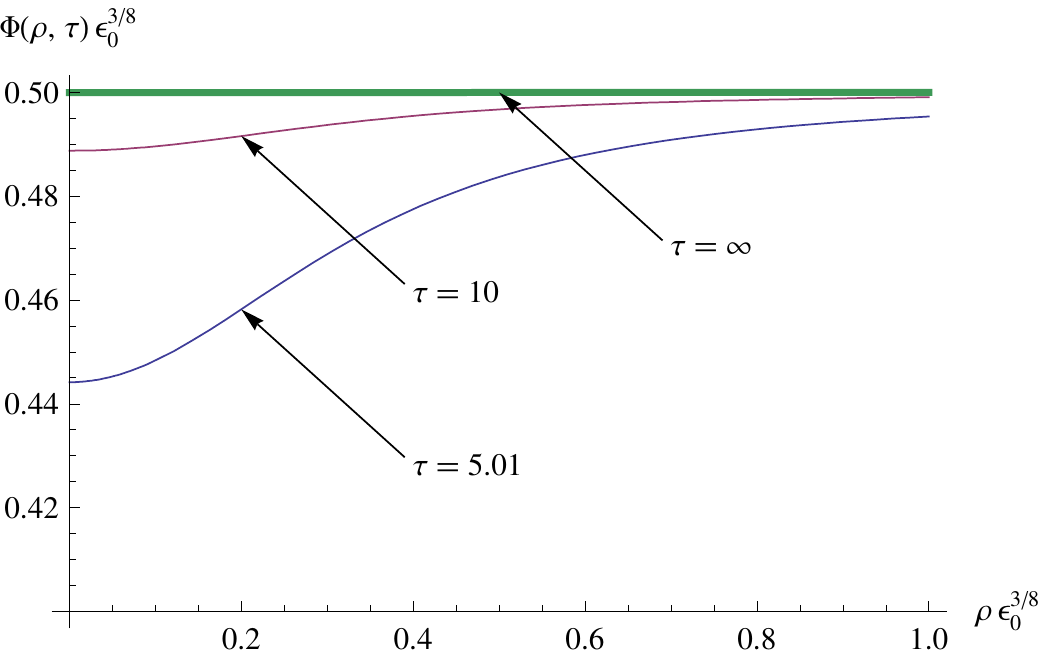}
\caption{\label{fig.embeddings}Embeddings with $m=\frac{1}{2}
  \varepsilon^{-3/8}$ at different times. For late times (bold), the
  supersymmetric embedding is approximated. The diagonal line in the
  first plot corresponds to $\Phi(r)\equiv r$, where the embeddings
  have to end. }
\end{figure}

The equation of motion arising from \eqref{d7embedaction} is a
non-linear partial differential equation. We will solve it
perturbatively by a late-time expansion
\begin{align}
  \Phi(\rho,\tau) &= m + \sum_{i=1}^\infty f_i(\rho) \tau^{-\frac{i}{3}}.
\end{align}
We use a fraction of $1/3$ in the exponent because all exponents
showing up in the background geometry \eqref{e.abcmetric} are integer
multiples of one third. The ansatz reduces the equations of motion to
the following (infinite) system of ordinary differential equations
\begin{gather}
   \rho^{-3} \p_\rho (\rho^3 f'_i(\rho) ) = \mathcal{I}_i(\rho) \label{embedode} \\
  \mathcal{I}_i = \frac{8 m \varepsilon_0^2 }{9(m^2 + \rho^2)^5} \cdot \begin{cases}
   1  & \text{if $i=8$}\\
   -4 \eta_0 \varepsilon_0^{-1/4} & \text{if $i=11$}\\
   0  & \text{else; provided $i<14$}
  \end{cases}\notag
\end{gather}
The boundary behavior of solutions to \eqref{embedode} is
\begin{align}
  f_i(\rho) \xrightarrow[\rho\to\infty]{} m_i + \frac{c_i}{\rho^2},
\end{align}
which becomes an exact solution when the inhomogeneous term vanishes,
$\mathcal{I}_i=0$.  The first term, $m_i$, contributes $m_i
\tau^{-i/3}$ to the bare quark mass.  Since we do not accept a
time-dependence of the bare parameters on physical grounds, we require
$m_i=0$.  Thus in conjunction with regularity the value of $c_i$ is
completely fixed.  In particular $\mathcal{I}_i=0$ implies $c_i=0$ or
$f_i\equiv0$.

\begin{figure}
\includegraphics[width=\figwidth]{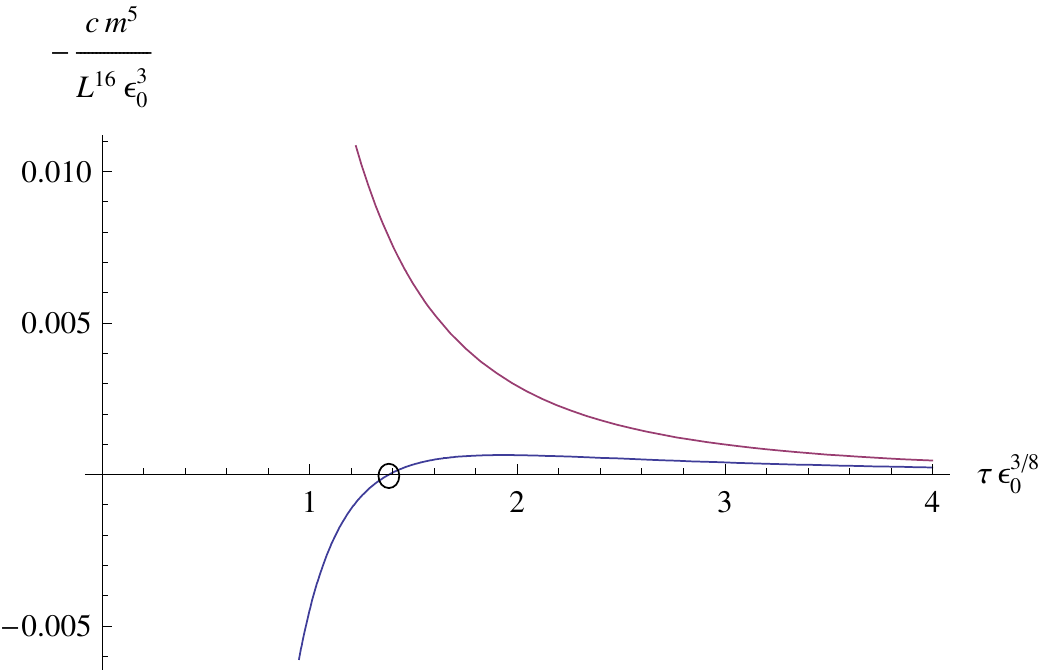}
\caption{\label{fig.cond}Time-dependence of the chiral condensate for
  adiabatic approximation (monotonic curve) and with viscosity
  correction. The circle at $\tau=(4/3)^{9/8}\approx 1.38$ is
  outside the regime of validity given by equation \eqref{e.valid}.
}
\end{figure}

To the considered order the solution is
\begin{align} \label{eq:d7sol}
  \Phi(\rho,\tau) &= m + c \,
     \frac{\rho^4 + 3\rho^2m^2 + 3m^4}{(m^2+\rho ^2)^3},
\end{align}
with
\begin{align}
  c = - \frac{\varepsilon_0^2L^{16}}{54m^5} \tau^{-\frac{8}{3}} \left( 1 - 4\eta_0 \varepsilon_0^{-\frac{1}{4}} \tau^{-\frac{2}{3}} + \dots \right).
\end{align}
In figure \ref{fig.cond} we show the effect of the subleading term on
the shape of $c(\tau)$. Note that we can trust this result only as
long as this term is small compared to the leading order,
\begin{align} \label{e.valid}
  4 \eta_0 \varepsilon_0^{-1/4} \tau^{-2/3} &\ll 1. 
\end{align}

We observe that the viscosity term enters the chiral condensate
exactly the same way as it enters the energy density squared. We may
therefore express the chiral condensate as\footnote{We would like
  remind the reader that $\varepsilon(\tau)\sim T^4 N_c^2$, such that
  $\vev{\cO}\sim N_c N_f T^8$.}
\begin{align} \label{e.condeps}
  \vev{\cO} = \frac{1}{216\pi^4} \frac{N_f \lambda^3}{N_c^3} \frac{\varepsilon^2}{m_q^5}.
\end{align}

This solution is however only valid for late times as each term in
$\varepsilon(\tau)$ apparently diverges for $\tau\to0$.  While at this
stage it is not clear up to which time (if at all) our perturbative
expansion converges, we may still ask for which range of $\tau$ it is
self-consistent. For this consideration, three time-scales are of
potential importance. Firstly, the time when the solution touches the
horizon, i.e., $\Phi(0,\tau)= 3^{-1/4}L^2 \varepsilon_0^{1/3}
\tau^{-1/3}$.  Secondly, the moment when the viscosity term in the
expansion dominates the leading term in such a way that the embedding
``recoils'' and stops being a one-valued function of the holographic
direction $r$. This happens when $d^2\Phi(r,\tau)/dr^2\to\infty$.
Thirdly, the time $\tau=(4/3)^{9/8}\approx 1.38$ when the chiral
condensate changes its sign and by \eqref{e.condeps} would lead to an
imaginary energy density.
Before this time scale is reached, the regime of validity \eqref{e.valid}
is left.

\begin{figure}
\includegraphics[width=\figwidth]{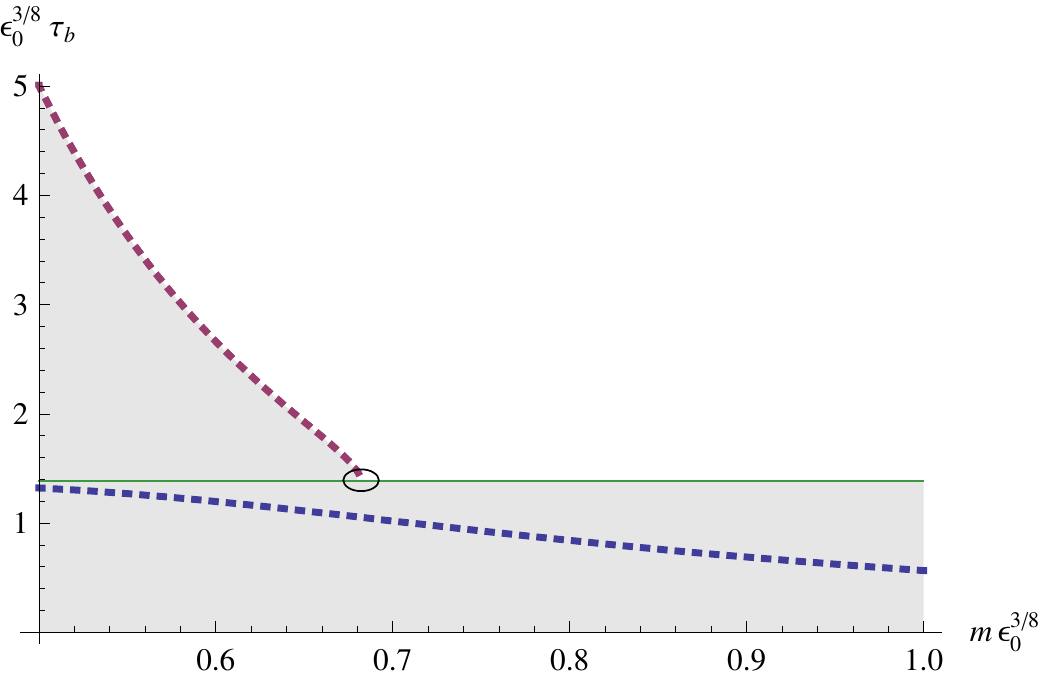}
\caption{\label{fig.breakdowntimes}Break-down time of the solution as
  a function of the quark mass. The horizontal line indicates the time
  where the energy density becomes imaginary, the dashed flat curve 
  is where embeddings become two-valued as a function of the holographic
  energy scale $r$ and the steep curve shows where solutions touch the
  horizon, i.e., can no longer be considered `Minkowski-type'. 
  The gray area marks invalid solutions. The circle is at $m\approx 0.682\varepsilon_0^{-3/8}$,
  which is the smallest mass that does not lead to solutions that eventually touch
  the horizon.}
\end{figure}

For numerical computations and plots, we will use
$\varepsilon_0^{-3/8}$ (which is a length) as a unit to express
dimensionful quantities.  Figure~\ref{fig.embeddings} shows the $m=
1/2\varepsilon_0^{-3/8}$ solution at various times before 
break-down. (Again we would like to stress that the solutions may
be invalid even before -- by ``break-down'' we denote their having
become invalid for sure.)  Figure~\ref{fig.breakdowntimes} compares the three
effects. For small quark mass solutions are invalidated by touching
the horizon, and by imaginary energy density for large quark mass.

\section{Meson spectra}
In the D3/D7 framework, meson spectra are determined from regular,
normalizable solutions to the equations obtained from linearizing the
full equations of motion of the D7 brane about the embedding solution
that describes the position and shape of the brane \cite{KMMW}.

In the following section we distinguish between four dimensional meson
modes, which carry a ``4d'' label and eight dimensional fluctuations,
which always start with a $\delta$ followed by a (Greek or Latin)
capital letter, e.g. $\delta\Phi$ or $\delta A^y$. Our ans\"atze are
products of spherical harmonics $\cY$ on the internal manifold, wave
forms parallel to the boundary and radial parts, which describe the
dependence on the holographic coordinates and are denoted by $\delta$
followed by a small letter.

\subsection{Boost invariance for the AdS geometry}
Before turning to the actual holographic computation of meson spectra
for the viscous fluid, it is insightful to investigate how boost
invariance changes wave forms of scalar and vector mesons in the
conventional setting, where there is no time-dependence of the meson
mass.

In four dimensional Minkowski space the massive Klein--Gordon equation assumes the form
\begin{gather}
  \Box \Phi_{4d}
    = \left[ -\frac{1}{\tau} \p_\tau \tau \p_\tau + \tau^{-2} \p_y^2 + \p_x^2 \right] \Phi_{4d} = M^2 \Phi_{4d} \\
  \implies \Phi_{4d} = ( c_1 \BesselJ0(\omega\,\tau) + c_2 \BesselY0(\omega\,\tau) ) e^{\pm ik_\perp x_\perp}  \\
  k_\perp^2 = k_2^2 + k_3^2 \notag
\end{gather}
with $\BesselJ0$ and $\BesselY0$ Bessel functions of first kind.
Since a linear combination of the Bessel functions will appear
frequently in our expressions we introduce the short hand
\begin{align} \label{e.Fdef}
  \cF_p[\omega] := c_1 \BesselJ{p}(\tint \omega \, d\tau) + c_2 \BesselY{p}(\tint \omega \,d\tau).
\end{align}
We will not explicitly denote the time-dependence of $\cF_0$ arising
from the integral over $\tau$.  At this stage, the integral has been
chosen for later convenience and gives $\omega\, \tau$ for constant
frequencies.  The eigenfrequencies $\omega := \sqrt{M^2 + k_\perp^2}$
are to be determined in our holographic setup.  We will thus assume
$k_\perp=0$ from the start to obtain the mass spectrum.

The 4d meson field is given as the boundary value of (linear,
normalizable) fluctuations $\delta\Phi$, $\delta\Psi$ about the
embedding solution $\Phi(\rho)\equiv m$,
\begin{align} \label{e.x8x9}
  X^8 &= 0 + \delta \Psi(\rho,\tau), &
  X^9 &= \Phi(\rho) + \delta\Phi(\rho,\tau).
\end{align}
For the presentation of our ansatz, we will concentrate on the scalar
mode $\delta\Phi$; the pseudoscalar mode $\delta\Psi$ will be treated
analogously.  With the following holographic ansatz
\begin{align} \label{e.holostatic}
  \delta\Phi(\rho,\tau) = \delta\phi(\rho) \cF_0[\omega]  \cY^\ell(S^3),
\end{align}
the boundary value corresponding to quantum number $\ell$ is defined
by
\begin{align} \label{e.bulkboundary}
  \Phi_{4d}^{(\ell)} = \lim_{\rho\to\infty} \rho^2 \frac{\delta\Phi(\rho,\tau)}{ \cY^\ell(S^3) }
\end{align}
Equation \eqref{e.holostatic} is a natural modification of the ansatz
given in \cite{KMMW} to separate the D7 equation of motion in
anti-de~Sitter space:
\begin{align}
  \left[ - \frac{L^2}{(\rho^2 + m^2)^2} \frac{1}{\tau} \p_\tau \tau \p_\tau +
  \frac{1}{\rho^3} \p_\rho \rho^3 \p_\rho + \frac{1}{\rho^2}
  \Delta_{S^3} \right] \delta\Phi(\rho,\tau) = 0.
\end{align}
The radial equation obtained after separation reads
\begin{align}
  \left[
  \frac{1}{\rho^3} \p_\rho \rho^3 \p_\rho +
  \frac{L^2\omega^2}{\rho^2 + m^2}
  - \frac{\ell(\ell+2)}{\rho^2} \right] \delta\phi(\rho) = 0,
\end{align}
which is the well-known result of \cite{KMMW}.  For simplicity, we
will only consider the lowest Kaluza--Klein mode on the internal
$S^3$, such that $\ell=0$, $\cY^0\equiv1$.

The requirements of regularity in the interior ($\rho \to 0$) and
vanishing at the boundary ($\rho \to \infty$), fix the modes
completely. One obtains a discrete set of modes the lightest of which
is given by
\begin{equation}
\label{e.solads}
\cF_0\left[\omega_0=\frac{\sqrt{8}m}{L^2} \right] \cdot \frac{1}{m^2+\rho^2}
\end{equation}

%
%

For a four-dimensional massive vector meson we have
\begin{align} \label{e.vec}
  \nabla_a F^{ab} &= M^2 A_{4d}^b.
\end{align}
We assume that the solutions are still plane waves in the
$x_2,x_3$\NB-plane, i.e., $A_{4d}^a=\xi^a(\tau)\exp ik_\perp x_\perp$.
This yields the following component equations
\begin{align}
  - \tau \p_\tau  ( \frac{1}{\tau} \p_\tau A^{4d}_y ) &= (M^2 + k_\perp^2 ) A^{4d}_y \label{minkowskiY} \\
  - \p_\tau^2 A^{4d}_2  - \frac{1}{\tau} \p_\tau A^{4d}_2  + i k_2 & \Bigl( \p_\tau  + \frac{1}{\tau} \Bigr)
       A^{4d}_\tau  \label{minkowski3} \\ + k_2 k_3  A^{4d}_3 & = (M^2+k_3^2) A^{4d}_2  \qquad \text{and ($2\leftrightarrow3$)} \notag \\
  A^{4d}_\tau &= -\frac{i \p_\tau ( k_2 A^{4d}_2 + k_3 A^{4d}_3 ) }{ \omega^2 } \label{minkowskiT}
\end{align}
Equation \eqref{minkowskiY} can be treated separately. Its solution is
\begin{align}
  A^{4d}_y &= \tau \cF_1[\omega] e^{ik_\perp x_\perp} &
  A^{4d}_{\tau,x^2,x^3} &= 0.
\end{align}

The others may be solved without loss of generality by turning the
coordinate system such that $k_3=0$. Then it follows immediately that
$A^{4d}_{2,3} = \xi_{2,3} \cF_0[\omega_{2,3} ] \exp ik_\perp x_\perp$.
With this modified ansatz and plugging \eqref{minkowskiT} into
\eqref{minkowski3} we obtain
\begin{align}
  \biggl[ -M^2 - k_3^2 + \omega_2^2 \left(1-\frac{k_2^2}{\omega^2}\right) \biggr] A^{4d}_2 + k_2 k_3 \biggl(1-\frac{\omega_3^2}{\omega^2}  \biggr) A^{4d}_3 &= 0 \notag \\ 
  \text{and ($2\leftrightarrow3$)},& 
\end{align}
which can only be satisfied for $\omega_2=\omega_3=:\omega_{23}$.
Moreover, since it is a homogeneous system, $\omega_{23}(M,k_2,k_3)$
can be determined from degeneracy of the coefficient matrix.  We shall
not reproduce the final expression, but just note that $\omega_{23} =
M$ when $k_2=k_3=0$, which could also have been obtained directly from
\eqref{minkowski3}.  We thus end up with the two solutions
\begin{align}
  A^{4d}_{2,3} = \xi_{2,3} \cF_0[\omega_{23}] \exp ik_\perp x_\perp.
\end{align}

Therefore, we adapt the holographic ans\"atze for meson modes found in
\cite{KMMW} as follows
\begin{align}
  \rlap{Type} \notag \\
  &\text{I}  &\delta A_\alpha &= \delta a_I^\pm(\rho) \cF_0[\omega] e^{ik_\perp x_\perp}
                                \cY^{\ell,\pm}_\alpha(S^3), \quad\!\! \alpha=5,6,7; \notag \\
  &\text{II}_y &\delta A_y &= \delta a_{IIy}(\rho) \tau\cF_1[\omega] e^{ik_\perp x_\perp}
                              \cY^{\ell}(S^3); \notag \\
  &\text{II}_{2,3}\!\!\! &\delta A_2 &= \delta a_{II2}(\rho) \cF_0[\omega] e^{ik_\perp x_\perp} 
                              \cY^{\ell}(S^3),  \quad  \!\! A_3 =0; \nonumber \\
              &                 &     & \qquad\qquad \text{and $(2\leftrightarrow 3)$} \notag \\
  &\text{III} &\delta A_\rho &= \delta a_{III}(\rho) \cF_0[\omega] e^{ik_\perp x_\perp}
                               \cY^{\ell}(S^3), \notag \\
                  &&\delta A_\alpha&= \delta \tilde a_{III}(\rho) \cF_0[\omega] e^{ik_\perp x_\perp}
                               \cY^{\ell,\pm}_\alpha(S^3);
\end{align}
with the respective other components set to zero.  We will only
consider modes of type II, which are the only modes dual to vector
mesons and therefore most interesting.

In the ans\"atze, the dependence in the $0,1,2,3$ directions has been
modified as compared to what can be found in \cite{KMMW}.  The reason
these changes are straight-forward is the following: The calculation
of \cite{KMMW} only uses two important properties of the ans\"atze
regarding derivatives in those directions,
\begin{align}
  \Delta_{4d} \delta A_I &= M^2 \delta A_I, \qquad I \in [0,\dots,7] \label{e.delta} \\
  g_{4d}^{ab} \p_a \delta A_b &= 0, \label{e.gauge}
\end{align}
where $g_{4d} = \diag (-1 , \tau^2 , 1, 1)$ in our case, whereas in
\cite{KMMW} it was a Minkowski metric.  For our ans\"atze, the gauge
condition \eqref{e.gauge} is either trivially obeyed or follows from
\eqref{minkowskiT}. Moreover it can be used to turn \eqref{e.delta}
into \eqref{e.vec}.

\subsection{Viscous fluid geometry}

Before coming to the actual holographic computation, we 
would like to discuss the general framework of late-time perturbative
expansions that we use.

Since the viscous fluid geometry and our
D7 embeddings are time-dependent, we do not expect, and do not see, a
separation into a purely $\tau$ dependent and $\rho$ dependent
factor. This makes the problem very difficult to tackle
analytically. We are helped by the property that at late proper times
the geometry becomes pure $AdS_5$ with the corresponding D7 brane
embedding. In this limit the simplest solution looks like (\ref{e.solads})
\begin{equation}
\label{e.soladsii}
\cF_0\left[\omega_0=\frac{\sqrt{8}m}{L^2} \right] \cdot \frac{1}{m^2+\rho^2},
\end{equation}
where $\cF_0[\omega]$ is defined in equation \eqref{e.Fdef}.
For smaller proper-times it is natural to treat the frequency
appearing in (\ref{e.soladsii}) as depending on $\tau$. However as the
equations do not allow for a separation of variables we have $\tau$
dependence also in the remaining part:
\begin{equation}
\label{e.split}
\tilde{\cF}\Bigl[\omega(\tau)\Bigr] f(\rho,\tau)
\end{equation}
where we allow for a general $\tilde{\cF}$ which should reduce to
$\cF_0[\omega]$ for {\em constant} $\omega$. We have moreover
the expansions
\begin{align}
\omega(\tau) &= \omega+\frac{1}{\tau^{\frac{1}{3}}} \omega^{(1)}+\ldots \\
f(\rho,\tau) &= f^{(0)}(\rho)+\frac{1}{\tau^{\frac{1}{3}}} f^{(1)}(\rho)+\ldots 
\end{align}
Note that the above form is not unique. Redefining the coefficients of
the expansions in an appropriate way, we may redefine the split
(\ref{e.split}). So in order to uniquely specify such an ansatz we
have to supplement the usual regularity condition at $\rho=0$ and
Dirichlet boundary condition at $\rho=\infty$ by another condition
which makes the split (\ref{e.split}) unique. In this paper we will
impose a condition on the profile of the mode $\delta \phi$ induced on
the boundary
\begin{equation} \label{e.induced}
\Phi_{4d}(\tau) \equiv \lim_{\rho \to \infty} \rho^2 \delta \phi(\rho,\tau)
\end{equation}
Namely we will set 
\begin{align}
  \Phi_{4d}(\tau) &= \sqrt{\frac{ \tint \omega_{4d}(\tau)\,
      d\tau}{\omega_{4d}\,\tau}}  
\cF_0[\omega_{4d}(\tau)]. \label{e.WKBpre}
\end{align}
which provides a definition of our frequency $\omega_{4d}(\tau)$. For
constant $\omega_{4d}(\tau)$ this reduces of course to the pure
$AdS_5$ result \eqref{e.soladsii}.

Our motivation for the above form (\ref{e.WKBpre}) is that it arises
as a \acro{WKB} approximation to a Klein--Gordon equation with time dependent
mass spectra: 
\begin{align} 
 \Box \Phi_{4d} &
    = \left[ -\frac{1}{\tau} \p_\tau \tau \p_\tau + \tau^{-2} \p_y^2 + \p_x^2 \right] \Phi_{4d} = M^2_{4d}(\tau) \Phi_{4d} 
\end{align}
We may separate variables by assuming a plane wave in the $2,3$
plane and obtain $\omega_{4d}^2(\tau)=M_{4d}^2(\tau) +
k_\perp^2$. (Though we will assume $k_\perp =0$, henceforth.) The
remaining equation
\begin{align} \label{e.timedep.KG}
  - \frac{1}{\tau} \p_\tau \tau \p_\tau  \Phi_{4d}(\tau) = \omega^2_{4d}(\tau) \Phi_{4d}(\tau)
\end{align}
can only be solved approximately, e.g., by the \acro{WKB} approximation, which 
gives two linearly independent solutions,
\begin{align}
  \Phi_{4d}(\tau) &\approx \sqrt{\frac{ \tint \omega_{4d}(\tau)\, d\tau}{\omega_{4d}\,\tau}} 
\cF_0[\omega_{4d}(\tau)]. \label{e.WKB}
\end{align}
The square root prefactor ensures that Abel's theorem is fulfilled, such
that the Wronskian for our ansatz is 
\begin{align} \label{e.Wronskian}
  W = \frac{\text{const}}{t}
\end{align}
as it should be for the exact solution.


We have now all ingredients in place to actually calculate the meson
spectrum for the time-dependent viscous fluid geometry.  We expand the
D7 action \eqref{d7action},\footnote{We do not write out those quartic terms
  that can only produce terms quartic in fluctuations.} given by
\begin{align}
 \Lag_{DBI} & = e^{\frac{\mathcal{A}}{2}+\frac{\mathcal{B}}{2}+\mathcal{C}} \rho ^3 \tau \biggl[
   1 + (\p_\rho X^9)^2+(\p_\rho X^8)^2+e^{-\mathcal{C}} (\p_\rho A_2){}^2 \notag \\
   & \qquad \qquad +\frac{e^{-\mathcal{B}}}{\tau^2} (\p_\rho A_y){}^2 -
\frac{e^{-\mathcal{A}}L^4}{r^4} (\text{I}) + \text{(quartic)} \biggr]^{1/2}  \notag \\[2ex]
(\text{I})  & =  
          (\p_\tau X^8)^2 
         + (\p_\tau X^9)^2
         + (\p_\rho X^8)^2 (\p_\tau X^9)^2 \notag \\ & \qquad
         + (\p_\tau X^8)^2 (\p_\rho X^9)^2 \notag \\ & \qquad
         - 2 (\p_\tau X^8) (\p_\rho X^9) (\p_\rho X^8) (\p_\tau X^9)
         \notag \\ &\qquad
         + e^{-\mathcal{C}} (\text{II}) 
         + \frac{e^{-\mathcal{B}}}{\tau ^2} (\text{III}) \\
(\text{II}) &= (\p_\tau A_2){}^2 
        + (\p_\rho X^9)^2 (\p_\tau A_2){}^2
        + (\p_\tau X^9)^2 (\p_\rho A_2){}^2 
\notag \\ & \qquad
        - 2 (\p_\tau X^9) (\p_\rho X^9) (\p_\rho A_2) (\p_\tau A_2) \\
(\text{III}) &=
         + (\p_\tau A_y){}^2 + (\p_\rho X^9)^2 (\p_\tau A_y){}^2 
         + (\p_\tau X^9)^2 (\p_\rho A_y){}^2 \notag \\ & \qquad
         - 2 (\p_\tau X^9) (\p_\rho X^9) (\p_\rho A_y) (\p_\tau A_y) 
\end{align}
to quadratic order in fluctuations
\begin{align}
  X^9 &= \Phi + \delta \Phi, &
  X^8 &=  0   + \delta \Psi, \notag \\
  A_2  &=  0   + \delta A_2, &
  A_y  &=  0   + \delta A_y.
\end{align}
The resulting equation of motion is evaluated by performing a
perturbative expansion in $\tau^{-1/3}$,
\begin{align} \label{e.holoansatz}
  \delta\Phi &= c_1 \BesselJ0 \Bigl( \tint \omega^{(\phi)}(\tau) \,d\tau \Bigr) \sum_{j=0}^\infty \delta\phi_j(\rho) \, \tau^{-j/3} \notag \\ \quad&
              + c_2 \BesselY0 \Bigl( \tint \omega^{(\phi)}(\tau) \,d\tau \Bigr) \sum_{j=0}^\infty \delta\tilde\phi_j(\rho) \, \tau^{-j/3}, \\
  \omega^{(\phi)} &= \sum_{i=0}^\infty \omega_i \, \tau^{-\frac{i}{3}},
\end{align}
and analogously for the other fluctuations.\footnote{The equation of
motion of $A_y$ requires a slightly modified ansatz given in the appendix.}
We use the known
asymptotic expansion of the Bessel functions
%
and obtain schematically the following equation
\begin{align}
 &  [ \text{polynomial in $\tau^{-1/3}$} ] \cos \bigl( \tint \omega^{(\phi)}\,d\tau\bigr) \notag \\ & \qquad \qquad 
+ [ \text{polynomial in $\tau^{-1/3}$} ]\sin \bigl( \tint \omega^{(\phi)}\,d\tau \bigr) = 0.
\end{align}
At any given order, the requirement that the coefficients of the
polynomials vanish, provides a differential equation for
$\delta\phi_i$ and $\delta\tilde\phi_i$ depending on $\omega_i$ (and
lower order solutions).  We have to go to order 6 before the viscosity
$\eta_0$ enters the equations. To this order, the equations for
$\delta\phi_i$ and $\delta\tilde\phi_i$ can be separated by choosing
suitable linear combinations and yield
$\delta\phi_i\equiv\delta\tilde\phi_i$, which is what is required
for the \acro{WKB} ansatz \eqref{e.WKB} to be applicable. 
We impose the boundary
conditions
\begin{gather}
  \delta\phi \xrightarrow[\rho\to\infty]{} 0, \qquad
  \delta\phi \xrightarrow[\rho\to0]{} \text{finite}, \notag \\
  \rho^2 \delta\phi \xrightarrow[\rho\to\infty]{} \sqrt{\frac{\tint \omega^{(\phi)}\, d\tau}{\omega^{(\phi)} \tau} } \cF_0[\omega^{(\phi)}].
\end{gather}
The first two of these conditions pick regular normalizable solutions;
the last ensures that meson solutions on the boundary \eqref{e.induced} satisfy the
constraint \eqref{e.Wronskian}. 
Consequently, the conditions fix two integration constants and the frequency
$\omega_i$ at each order in the perturbative expansion. 
The only remaining free constants are the overall factors $c_1$ and $c_2$
of our ansatz \eqref{e.holoansatz}.

Each of the coefficient functions has to satisfy a differential
equation that is best expressed with the substitutions
\begin{align}
  \delta\phi_i(\rho) &= (1-\bfy)^{-n-1} \delta\phi_i(\bfy),  \notag \\
  \bfy &= -\rho^2/m^2,  \\
  \omega_0^2 &= m^2 ((2n+3)^2-1). \notag
\end{align}
The lowest order equation then reads
\begin{equation} \label{mesonhyper}
\begin{gathered}
  \left[\bfy(1-\bfy)\partial_\bfy^2 + ( \bfC - (\bfA+\bfB-1) \bfy ) \partial_\bfy - \bfA \,\bfB \right] \phi_0(\bfy) = 0 \\
  \bfA = -n-1, \qquad
  \bfB = -n, \qquad
  \bfC = 2.
\end{gathered}
\end{equation}
This is exactly the hypergeometric equation already encountered in
\cite{KMMW}. The boundary conditions fix $n$ to be a non-negative
integer, thus yielding a discrete meson spectrum $M=\omega_0(n)$ and
the solutions in terms of (degenerate) hypergeometric functions $\F$
are
\begin{align}
  h_1 &= \F( -n -1 , -n, 2; \bfy), \\
  h_2 &= (1-\bfy)^{3+2n} \F( n + 2, n +3, 2n+4; 1-\bfy).
\end{align}
Only $h_1$ is regular, such that $\delta\phi_0\equiv h_1$.

\begin{figure}
\includegraphics[width=\figwidth]{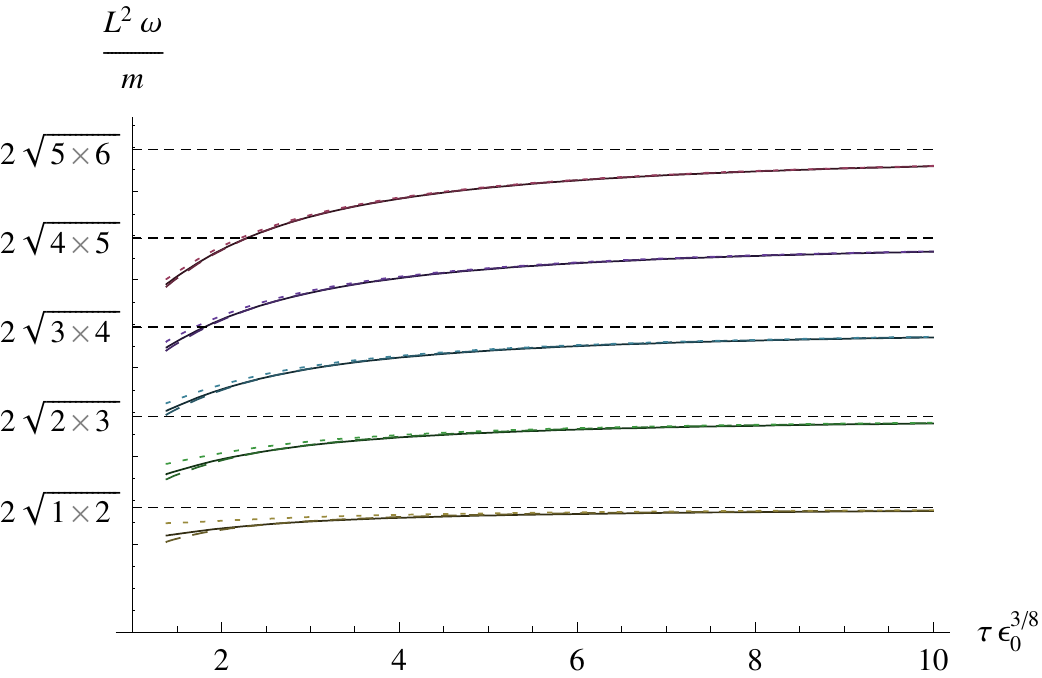}
\caption{\label{fig.spectrum}Late-time spectra for the viscous fluid
  geometry. The supersymmetric spectrum is shown as dashed horizontal
  lines. Scalar mesons are shown as continuous lines, vector modes are
  dashed.}
\end{figure}

Higher orders in perturbation theory produce inhomogeneous terms in
the analogues of \eqref{mesonhyper}. Since it is a linear ordinary
equation, the solution can still be obtained in closed form by standard
methods.  However, the resulting integrals are hard to solve in
general. Since both solutions $h_{1,2}$ are rational functions of
$\bfy$ and $\ln \bfy$, it is however easy to do so for definite
$n$. For the lowest five mesons $n=0,\dots,4$ we give the solutions in
the appendix.

The mass\footnote{defined by equation \eqref{e.WKBpre}} 
of the lowest scalar meson mode
is
\begin{align}\label{e.M}
  \omega^{(\phi)} 
  =
\tfrac{4 \pi }{\sqrt{\lambda }} \cdot \left[ m_q - \tfrac{3 \lambda ^2 \varepsilon _0}{80 \pi ^4 \tau ^{4/3} m_q^3}\cdot \left(1-\tfrac{2 \eta _0}{\tau^\frac{2}{3} \varepsilon_0^\frac{1}{4}}\right) \right].
\end{align}

To the considered order, pseudoscalar modes $\delta\psi$ have exactly
the same equations of motion and the spectrum is degenerate.  Moreover
we note that the spectrum agrees with the adiabatic approximation even
including the viscosity corrections. The reason for this might be that
the bulk metric coefficients that enter the calculation for the scalar
mesons can be expressed completely in terms of the energy density,
whereas the components for the $y,2,3$ directions cannot.

\begin{figure}
\includegraphics[width=\figwidth]{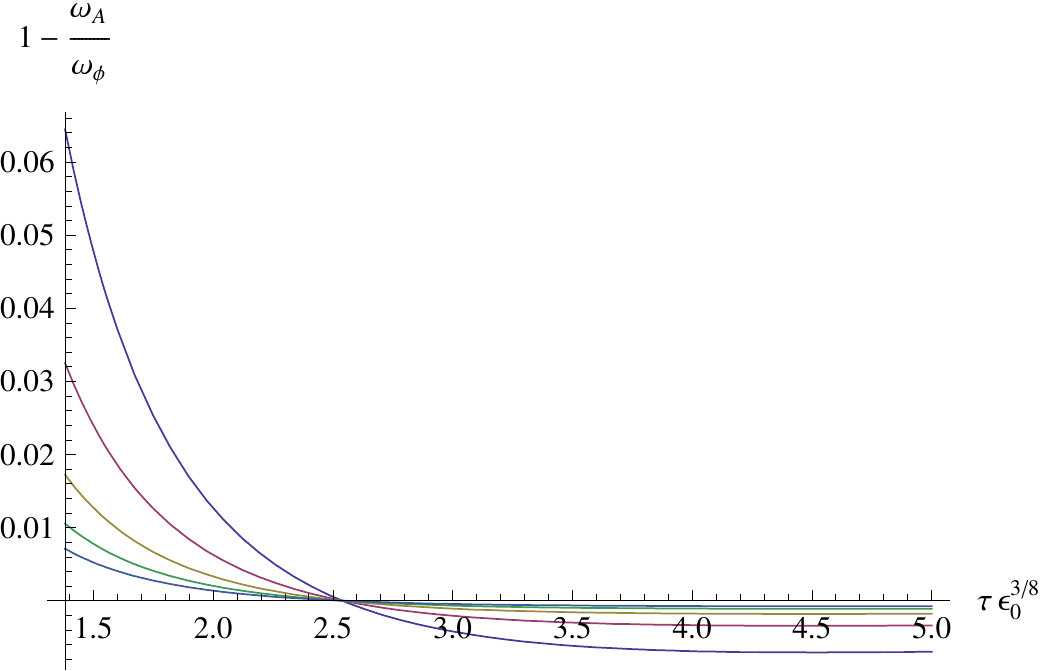}
\caption{\label{fig.cmpvectors} The plot shows the relative difference
  of the masses of (type II${}_y$) vector and scalar mesons for
  $m=\varepsilon_0^{3/8}$. }
\end{figure}

The vector mesons deviate slightly from the scalar modes.  For
comparison we plot the mass ratio of scalar and vector modes in
Figure~\ref{fig.cmpvectors}. The mass of the lowest vector mesons is
given by
\begin{align}
  \omega^{(A^y)}
  = \tfrac{4 \pi }{\sqrt{\lambda }} \cdot \left[ m_q - \tfrac{ 7 \lambda ^2 \varepsilon _0}{240 \pi ^4 \tau ^{4/3} m_q^3}\cdot \left(1-\tfrac{6 \eta _0}{7 \tau^\frac{2}{3} \varepsilon_0^\frac{1}{4}}\right) \right],\\
  \omega^{(A^{2,3})}
  = \tfrac{4 \pi }{\sqrt{\lambda }} \cdot \left[ m_q - \tfrac{ 7 \lambda ^2 \varepsilon _0}{240 \pi ^4 \tau ^{4/3} m_q^3}\cdot \left(1-\tfrac{18 \eta _0}{7 \tau^\frac{2}{3} \varepsilon_0^\frac{1}{4}}\right) \right].
\end{align}
This agrees with the adiabatic approximation excluding the viscosity term. 
Since the metric components that enter the holographic computation, $g_{yy}$ and $g_{22}$, agree only up to the viscosity terms, this deviation does not
come as a surprise.

We plot the five lowest meson modes in Figure~\ref{fig.spectrum}.
The leading order term gives the exact supersymmetric spectrum that is
approached for $\tau\to\infty$.

\subsection{Comparison to the adiabatic approximation}
In this subsection, we will review some properties of low-temperature
meson spectra for the static AdS black hole.  Plugging in the
time-dependence of the temperature into the static meson spectra,
yields an estimate for the time-dependent spectrum, which we will
refer to as \emph{adiabatic approximation}. We will assume that the
temperature dependence given in terms of \Poincare\ time $t$ can be
obtained from \eqref{e.T} by substituting $\tau$ for $t$:
\begin{equation} \label{e.Toft}
T_{AdS/BH}(t)=\left(\frac{4\varepsilon_0}{3}\right)^{\frac{1}{4}} \frac{1}{\pi\,  t^{\frac{1}{3}}} \left[ 1 - \frac{\eta_0}{2 \varepsilon_0^{1/4} t^{2/3}} \right].
\end{equation}
Note however that we do not expect the resulting adiabatic meson
spectra to accurately give the viscosity corrections. The reason for
this is that even though the horizon position can be expressed
completely in terms of the energy density, such that the
Stefan--Boltzmann law holds, the bulk metric and energy momentum
tensor nevertheless contain additional viscosity terms that cannot be
captured by the AdS/Schwarzschild geometry even when the geometry near
the horizon and near the boundary is matched.
\begin{figure}
\includegraphics[width=\figwidth]{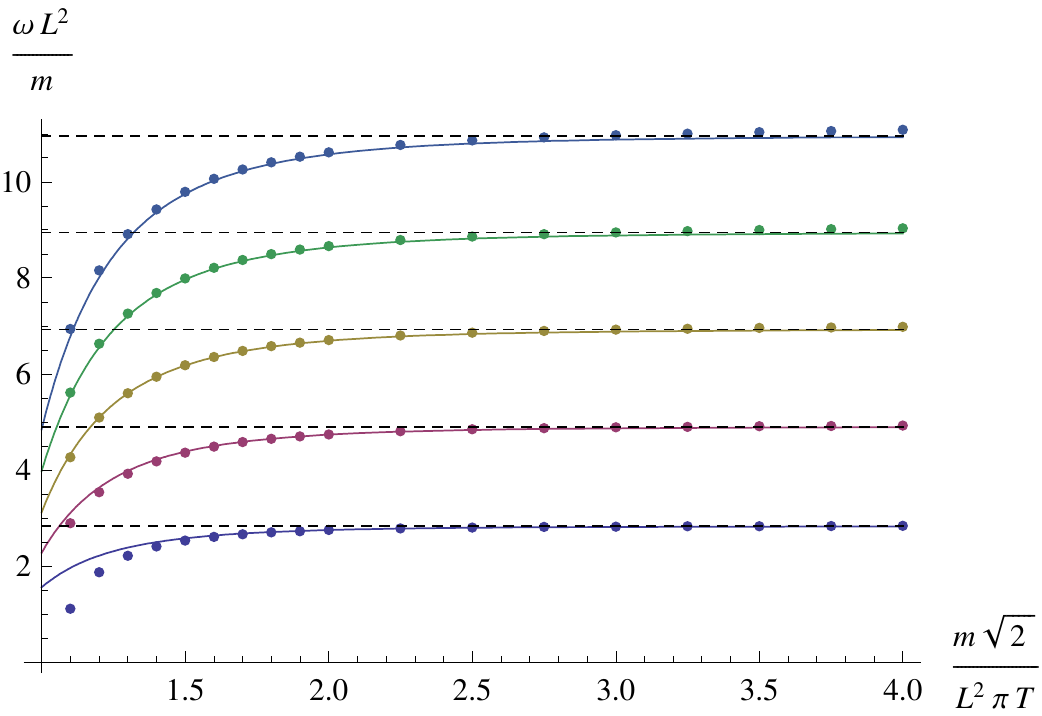}
\caption{\label{fig.staticspectrum}The plot shows the (pseudo)scalar
  meson spectrum. For small temperatures, the supersymmetric spectrum (dashed) 
  is approached. 
  The solid lines are asymptotic $T\to0$ solutions, which are in 
  good agreement with the numerical solutions (dots) for small temperatures.}
\end{figure}

In Figure~\ref{fig.staticspectrum} we plot the numerical solution
(dots) of the static case. We calculate the asymptotic solution in a
low temperature expansion (solid curves), which agrees with the
numerical calculation.  Plugging the time dependence of the
temperature \eqref{e.Toft} into this analytical approximation, we
obtain the time-dependent meson spectrum in adiabatic approximation.
The mass of the lowest scalar and vector modes are given by

\begin{align}
  \omega_{\phi,\psi}^{\text{ad.}} 
   &= \frac{4\pi}{\sqrt{\lambda}} \left[ m_q - \frac{9\lambda^2}{320} \frac{T^4}{m_q^3} \right] \notag \\
  &= \frac{4\pi}{\sqrt{\lambda}} \left[ m_q - \frac{3\lambda^2}{80\pi^4} \frac{\varepsilon_0}{m_q^3} t^{-\frac{4}{3}} 
\left(1-\tfrac{2 \eta _0}{t^\frac{2}{3} \varepsilon_0^\frac{1}{4}}\right)
\right],\\
  \omega_{A^{\mu}}^{\text{ad.}} 
   &= \frac{4\pi}{\sqrt{\lambda}} \left[ m_q - \frac{7\lambda^2}{320} \frac{T^4}{m_q^3} \right] \notag \\  
   &= \frac{4\pi}{\sqrt{\lambda}} \left[ m_q - \frac{7\lambda^2}{240\pi^4} \frac{\varepsilon_0}{m_q^3} t^{-\frac{4}{3}} 
\left(1-\tfrac{2 \eta _0}{t^\frac{2}{3} \varepsilon_0^\frac{1}{4}}\right)
\right].
\end{align}
The mass of all of these modes decreases for increasing
temperature.\footnote{A similar decrease of meson masses for
  increasing (static) temperature was found in the Sakai/Sugimoto
  model \cite{Kasper}.} Moreover, the adiabatic scalar modes
completely agree with our result \eqref{e.M}, whereas the vector modes
only agree in the leading contribution; the viscosity terms differ. We
consider the agreement of the scalars accidental in the sense that it
is a consequence of a certain property of the expanding plasma
geometry: All metric coefficients entering the scalar equation of
motion can be expressed purely in terms of the temperature, while
additional viscosity terms only show up in those metric coefficients
that end up in the equations for the vector modes.

An important assumption in our calculation has been that the
Klein--Gordon equation is obeyed by scalar particles.
This assumption can actually be proved employing the holographic
equation of motion resulting from the linearization procedure.
There is a parity symmetry $\rho \mapsto -\rho$ in the equation of
motion. Since only `Minkowski-type' solutions are considered, 
this will lead to even solutions, which have the following expansion,
\begin{align} \label{e.boundaryexpansion}
  \delta\Phi(\rho\to\infty,\tau) &\sim \Phi_{4d}(\tau) \frac{1}{\rho^2}  - \frac{L^4\Phi_{4d}(\tau)}{8} M^2(\tau) \frac{1}{\rho^4} + \dots,
\end{align}
because only normalizable solutions are allowed, such that the constant leading 
term vanishes. 
The subleading coefficient $M^2(\tau)$ has been multiplied by an additional factor $-L^4\Phi_{4d}/8$
for later convenience. Plugging above expansion into the eight-dimensional
equation of motion, yields at leading order in $1/\rho$,
\begin{align}
  - \frac{1}{\tau} \p_\tau ( \tau \p_\tau ) \Phi_{4d} &= M^2(\tau) \Phi_{4d}.
\end{align} 
This establishes that at least for a background geometry dual to a 
hydrodynamic expansion up to and including viscosity, the scalar 
meson equation is a Klein--Gordon equation with time-dependent mass. 
This might change when also encoding
higher order effects like the relaxation time \cite{Heller:2007qt},
which currently appears to be out of reach of a supergravity approximation
\cite{Benincasa:2007tp}.

We will now assess the error of the \acro{WKB} approximation
by plugging our meson solutions into the four dimensional Klein--Gordon
equation. The error should
be smaller than $\tau^{-2}$ to be subleading to the viscosity contribution.
We first determine the four dimensional meson solution by
\begin{align}
  \Phi_{4d}^{(\ell)} &= \lim_{\rho\to\infty} \rho^2 \delta\phi(\rho,\tau) \cF_0[ \omega^{(\phi)} ]
\label{e.4dholo}
\end{align}
where on the right hand side we plug in the mass \eqref{e.M}
of the lowest holographic scalar meson solution, i.e., we set $\omega_{4d} \approx \omega^{(\phi)}$.

With \eqref{e.4dholo} the error estimate $\Delta\omega$ can be obtained from 
the Klein--Gordon equation
\begin{align} 
  \frac{1}{\tau} \p_\tau \tau \p_\tau  \Phi_{4d}(\tau) = \bigl(\omega^{(\phi)}(\tau)+\Delta\omega(\tau)\bigr)^2 \Phi_{4d}(\tau) ,
\end{align}
by linearizing in $\Delta\omega$. This yields
\begin{align}
  \Delta\omega(\tau) = \frac{\sqrt{2}L^{10} \varepsilon_0}{5m^4} \frac{1}{\tau^{10/3}} + \dots ,
\end{align}
which is sufficiently small, that is subleading to the viscosity terms
arising from the geometry. (Also note that the frequencies
$\omega^{(\phi)}(\tau)$ and the meson mass obtained from the holographic
expansion \eqref{e.boundaryexpansion} agree up to and including order 
$\tau^{-2}$.)  However when encoding hydrodynamic effects in
the geometry that are of sufficiently high order, we would be forced
to consider a better approximation for our ansatz, e.g., by using
higher order \acro{WKB}. Moreover, beyond a certain order the
\acro{WKB} ansatz is expected not to work anymore because the
coefficients $\delta\phi_i$ and $\delta\tilde\phi$ are not expected to
coincide to arbitrary order.

\section{Regularized D7 action}
In the static case, the free energy density can be related to the
regularized D7 action by means of a Wick rotation. In the conventions
of \cite{Mateos:2007vn} the time direction is periodically identified
with $\beta=1/T$, such that the free energy density $F$ is given by
\begin{align} \label{freeenergydef}
  F &= T \cdot \frac{S_{D7}}{V_x},
\end{align}
with $T$ the temperature and $V_x=\int d^3x$ the (infinite) spatial
volume of the boundary.  The authors of \cite{Mateos:2007vn} obtain
\begin{align} \label{mateosmatter}
  \frac{S_{D7}}{V_x}
     &\sim - \frac{N_c N_f T^3 \lambda}{32} \cdot \frac{1}{12} \left(\frac{T}{2 m_q / \sqrt{\lambda}}\right)^4
      = -\frac{N_c N_f \lambda^3}{6114 m_q^4} \, T^7
\end{align}
in the limit of low temperature.  While it is clear that for a
time-dependent background neither $1/T =\int dt$ nor
\eqref{freeenergydef} really make sense, we are still interested in
how our action relates to above asymptotic result.

When calculating the action $S_{D7}$, the integration along the
holographic direction $\rho$ on the brane is encumbered by \acro{UV}
divergences.  The standard procedure of holographic renormalization
for the case of a probe D7 brane has been worked out in
\cite{Karch:2005ms}.  It consists in regularization by introducing a
cut off $\rho_{max}$ and addition of suitable counterterms formed from
the induced metric and corresponding curvature on the slice
$\rho=\rho_{max}$.
\begin{align}
  S_{reg}&=\int\limits_0^{\rho_{max}} \Lag_{D7}\,d\rho + \sum L_i(\rho_{max})
\end{align}
To be more precise, the procedure is usually formulated in terms of
the coordinate
\begin{align}
  z&=(\rho^2 + \Phi^2)^{-1/2}, \label{eq:zcoord}
\end{align}
where the counterterms $L_i$ read
\begin{align}
\begin{aligned}
  L_1 &= -\frac{1}{4} \sqrt{\gamma} \\
  L_2 &=-\frac{1}{48} \sqrt{\gamma} R_\gamma \\
  L_3 &= -\ln (z_{min}) \sqrt{\gamma} \frac{1}{32} ( R_{ij} R^{ij} - \frac{1}{3} R_\gamma ) \\
  L_4 &= \frac{1}{2} \sqrt{\gamma} \Psi^2 \\
  L_5 &= -\frac{1}{2} \ln (z_{min}) \Psi ( \p_\tau \gamma^{\tau\tau} \sqrt{\gamma} \p_\tau + \frac{1}{6} \sqrt{\gamma} R_\gamma ) \Psi \\
  L_f &= \alpha \gamma \Psi^4
\end{aligned}
\end{align}
$L_f$ is a finite counter term, with an arbitrary parameter $\alpha$,
that corresponds to different renormalization schemes. It can be fixed
in supersymmetric settings by the requirement that the free energy
vanishes.  $\gamma$ is the induced metric on the $z=z_{min}$ slice and
\begin{align}
\Psi&=\arcsin \left(\frac{z\Phi}{L^2}\right) \label{eq:psi}
\end{align}
is the embedding coordinate of the D7 expressed as an angle on the
internal $S^5$.

We express the counterterms in terms of $\rho_{max}:= (z_{min}^{-2}
-\Phi^2(\rho_{max}))^{-1/2}$ using equations \eqref{eq:psi} and
\eqref{eq:zcoord}
\begin{align}
  \sum L_i &= -\frac{1}{4} \tau \rho_{max}^4 + \tau m^4 \frac{5 + 12 \alpha}{12} + \OO{\rho_{max}^{-1}} \label{eq:countersum}
\end{align}
As a check we turn this back into $z$ coordinates by self-consistently
iterating \eqref{eq:zcoord}
\begin{align}
  \rho_{max}^2 &= \sqrt{z_{min}^{-2} + \Phi^2\left(\sqrt{z_{min}^{-2} + \Phi^2\left(\sqrt{z_{min}^{-2} + \Phi^2\left( \dots \right)} \right)} \right)}
\end{align}
and using the D7 embedding \eqref{eq:d7sol}. We obtain
\begin{align}
  \frac{1}{\tau} \sum L_i &= -\frac{1}{4} (m^2 - z_{min}^{-2})^2 - 4 m \, c
             + \frac{m^4(5+12 \alpha)}{12},
\end{align}
which up to the $\tau$-dependence arising from our rapidity
coordinates is exactly the result of \cite{Mateos:2007vn}, eq.~(4.21),
for $\alpha=-5/12$.

The renormalized action is thus
\begin{align}
   S_{ren} = \mathbf{N} \int d\tau \lim_{\rho_{max}\to\infty}\int\limits_0^{\rho_{max}} (\Lag_{DBI} - \rho^3 \tau)\,d\rho  \qquad & \notag \\[-1ex]
                                                                                       + \frac{1}{12} m^4 (5+12\alpha)\tau, &
\end{align}
where the last two terms are the divergent part of the counterterm,
suitably rewritten as an integral, and the remaining finite part. Of
course due to the renormalization, the limit can actually be carried
out. For late times we obtain
\begin{align}
  S_{ren} &= \mathbf{N} \int d\tau \,\tau  \left[
    \frac{m^4}{12} (12 \alpha+5)
    -\frac{\varepsilon_0^2 L^{16}}{108 m^4 \tau ^{8/3}}
    +\frac{\varepsilon_0^{7/4} \eta_0 L^{16}}{27 m^4 \tau ^{10/3}} \right]
\end{align}
such that for $\alpha=-5/12$ the configuration indeed relaxes to the
supersymmetric setting.

The above is our result. For comparison with \eqref{mateosmatter} we
perform an additional, somewhat ill-defined step, namely replacing
$\int d\tau\, \tau \mapsto 1/T $.  With $L^4 = 2 \lambda \ell_s^4$ and
equation \eqref{norm} we obtain for the supersymmetric scheme
\begin{align}
  S_{ren} &= - \frac{1}{3\cdot 2^{11}} \frac{N_c N_f \lambda^3}{m_q^4} \, T^7, 
\end{align}
which agrees with \eqref{mateosmatter}.

\section{Conclusions}
The main goal of this article was to study fundamental fields in the
holographic dual of an expanding viscous fluid. The dual geometry has
strong similarity to a black hole geometry with time-dependent
temperature. Small temperatures correspond to late-times.  We
determined the D7 embedding and calculated the consequences of
dynamical temperature for the chiral condensate to three orders.  The
leading order gives the supersymmetric solution, the subleading
corresponds to the adiabatic approximation and the subsubleading order
includes viscosity corrections going beyond the adiabatic
approximation.

Moreover we calculated the meson spectrum and find that it agrees 
in the subleading order, though only the scalar mesons 
agree in the subsubleading order with the adiabatic approximation.
The agreement crucially depends on the choice of 
ansatz defining the frequencies. 
We demonstrate that for our \acro{WKB} ansatz solves the 
Klein--Gordon up to an error smaller than that inescapably 
introduced in the late-time expansion of the geometry.

However, for the next order, which introduces the relaxation 
time \cite{Heller:2007qt}, higher order terms would be needed
in the \acro{WKB} approximation in order to still keep that
error smaller than the geometry's.

It would be interesting to consider analogous
properties for gauge theories which exhibit chiral symmetry breaking
and hence are more closer to \acro{QCD}.  However up till now there is
no description of an expanding plasma system in such a theory.

\section*{Acknowledgments}
\begin{sloppypar}
  We would like to thank Kasper Peeters and Marija Zamaklar for
  drawing our attention to static black hole meson spectra.  J.G.\
  acknowledges support by \acro{ENRAGE} (European Network on Random
  Geometry), a Marie Curie Research Training Network in the European
  Community's Sixth Framework Programme, network contract
  \acro{MRTN}-\acro{CT}-2004-005616. The work of P.S.\ was supported
  by a Jagiellonian University scholarship for graduate students.
  R.J.\ was supported in part by Polish Ministry of Science and
  Information Technologies grant \acro{1P03B04029} (2005-2008) and the
  Marie Curie \acro{ToK} project \acro{COCOS} (contract
  \acro{MTKD}-\acro{CT}-2004-517186).
\end{sloppypar}

\onecolumngrid

\appendix

\section{Meson solutions}
\subsection*{(Pseudo-)scalar mesons}
\begin{align*}
\delta \Phi _0&=\cF_0[\omega_0]\biggl[\tfrac{\left(8 m^4+9 \rho ^2 m^2+3 \rho ^4\right) \varepsilon _0 L^8}{10 m^4 \left(m^2+\rho ^2\right)^3 \tau ^{4/3}} 
	- \tfrac{\left(13 m^4+12 \rho ^2 m^2+3 \rho ^4\right) \varepsilon _0^{3/4} \eta _0 L^8}{10 m^4 \left(m^2+\rho ^2\right)^3 \tau ^2} 
	+ \tfrac{1}{m^2+\rho ^2}\biggr],\\ 
  \omega _0&=\tfrac{2 \sqrt{2} m}{L^2}+\left(-\tfrac{3 L^6 \tau^{-4/3} \varepsilon _0}{5 \sqrt{2} m^3}\right)\times \left(1-\tfrac{2 \eta _0}{\tau ^{2/3} \sqrt[4]{\varepsilon _0}}\right),  \displaybreak[0]\\[1.5ex] 
 \delta \Phi _1&=\cF_0[\omega_1]\biggl[\tfrac{\left(-12 m^6+15 \rho ^2 m^4+20 \rho ^4 m^2+5 \rho ^6\right) \varepsilon _0 L^8}{14 m^4 \left(m^2+\rho ^2\right)^4 \tau ^{4/3}}
	+ \tfrac{\left(19 m^6-35 \rho ^2 m^4-35 \rho ^4 m^2-5 \rho ^6\right) \varepsilon _0^{3/4} \eta _0 L^8}{14 m^4 \left(m^2+\rho ^2\right)^4 \tau ^2} 
	+ \tfrac{\rho ^2-m^2}{\left(m^2+\rho ^2\right)^2}\biggr],\\ 
  \omega _1&=\tfrac{2 \sqrt{6} m}{L^2}+\left(-\tfrac{5 \sqrt{3} L^6 \tau^{-4/3} \varepsilon _0}{7\sqrt{2} m^3}\right)\times \left(1-\tfrac{2 \eta _0}{\tau ^{2/3} \sqrt[4]{\varepsilon _0}}\right),  \displaybreak[0]\\[1.5ex] 
 \delta \Phi _2&=\cF_0[\omega_2]\biggl[\tfrac{\left(26 m^8-140 \rho ^2 m^6-22 \rho ^4 m^4+55 \rho ^6 m^2+11 \rho ^8\right) \varepsilon _0 L^8}{30 m^4 \left(m^2+\rho ^2\right)^5 \tau ^{4/3}} \\ 
&\qquad\qquad\qquad	- \tfrac{\left(41 m^8-269 \rho ^2 m^6+121 \rho ^6 m^2+11 \rho ^8\right) \varepsilon _0^{3/4} \eta _0 L^8}{30 m^4 \left(m^2+\rho ^2\right)^5 \tau ^2}
	+ \tfrac{m^4-3 \rho ^2 m^2+\rho ^4}{\left(m^2+\rho ^2\right)^3}\biggr],\\ 
  \omega _2&=\tfrac{4 \sqrt{3} m}{L^2}+\left(-\tfrac{11 L^6 \tau^{-4/3} \varepsilon _0}{5 \sqrt{3} m^3}\right)\times \left(1-\tfrac{2 \eta _0}{\tau ^{2/3} \sqrt[4]{\varepsilon _0}}\right),  \displaybreak[0]\\[1.5ex] 
 \delta \Phi _3&=\cF_0[\omega_3]\biggl[\tfrac{\left(-134 m^{10}+1506 \rho ^2 m^8-1650 \rho ^4 m^6-1045 \rho ^6 m^4+342 \rho ^8 m^2+57 \rho ^{10}\right) \varepsilon _0 L^8}{154 m^4 \left(m^2+\rho ^2\right)^6 \tau ^{4/3}} \\ 
&\qquad\qquad\qquad	+ \tfrac{\left(211 m^{10}-2784 \rho ^2 m^8+3585 \rho ^4 m^6+1805 \rho ^6 m^4-912 \rho ^8 m^2-57 \rho ^{10}\right) \varepsilon _0^{3/4} \eta _0 L^8}{154 m^4 \left(m^2+\rho ^2\right)^6 \tau ^2}\\ 
&\qquad\qquad\qquad	+ \tfrac{-m^6+6 \rho ^2 m^4-6 \rho ^4 m^2+\rho ^6}{\left(m^2+\rho ^2\right)^4}\biggr],\\ 
  \omega _3&=\tfrac{4 \sqrt{5} m}{L^2}+\left(-\tfrac{57 \sqrt{5} L^6 \tau^{-4/3} \varepsilon _0}{77 m^3}\right)\times \left(1-\tfrac{2 \eta _0}{\tau ^{2/3} \sqrt[4]{\varepsilon _0}}\right),  \displaybreak[0]\\[1.5ex] 
 \delta \Phi _4&=\cF_0[\omega_4]\biggl[\tfrac{\left(68 m^{12}-1279 \rho ^2 m^{10}+3543 \rho ^4 m^8-630 \rho ^6 m^6-1566 \rho ^8 m^4+203 \rho ^{10} m^2+29 \rho ^{12}\right) \varepsilon _0 L^8}{78 m^4 \left(m^2+\rho ^2\right)^7 \tau ^{4/3}} \\ 
&\qquad\qquad\qquad	- \tfrac{\left(107 m^{12}-2326 \rho ^2 m^{10}+7057 \rho ^4 m^8-1840 \rho ^6 m^6-3161 \rho ^8 m^4+638 \rho ^{10} m^2+29 \rho ^{12}\right) \varepsilon _0^{3/4} \eta _0 L^8}{78 m^4 \left(m^2+\rho ^2\right)^7 \tau ^2} \\ 
&\qquad\qquad\qquad	+ \tfrac{m^8-10 \rho ^2 m^6+20 \rho ^4 m^4-10 \rho ^6 m^2+\rho ^8}{\left(m^2+\rho ^2\right)^5}\biggr],\\ 
  \omega _4&=\tfrac{2 \sqrt{30} m}{L^2}+\left(-\tfrac{29 \sqrt{\tfrac{5}{6}} L^6 \tau^{-4/3} \varepsilon _0}{13 m^3}\right)\times \left(1-\tfrac{2 \eta _0}{\tau ^{2/3} \sqrt[4]{\varepsilon _0}}\right)
\end{align*}
\subsection*{Vector meson (Type \texorpdfstring{II${}_{2,3}$}{II(2,3)})}
\begin{align*}
(\delta a_{II2})_0&=\cF_0[\omega_0]\biggl[\tfrac{\left(22 m^4+21 \rho ^2 m^2+7 \rho ^4\right) \varepsilon _0 L^8}{30 m^4 \left(m^2+\rho ^2\right)^3 \tau ^{4/3}} 
	- \tfrac{\left(13 m^4+12 \rho ^2 m^2+3 \rho ^4\right) \varepsilon _0^{3/4} \eta _0 L^8}{10 m^4 \left(m^2+\rho ^2\right)^3 \tau ^2} 
	+ \tfrac{1}{m^2+\rho ^2}\biggr],\\ 
  \omega _0&=\tfrac{2 \sqrt{2} m}{L^2}+\left(-\tfrac{7 L^6 \tau^{-4/3} \varepsilon _0}{15 \sqrt{2} m^3}\right)\times \left(1-\tfrac{18 \eta _0}{7 \tau ^{2/3} \sqrt[4]{\varepsilon _0}}\right),  \displaybreak[0]\\[1.5ex] 
 (\delta a_{II2})_1&=\cF_0[\omega_1]\biggl[\tfrac{\left(-118 m^6+137 \rho ^2 m^4+164 \rho ^4 m^2+41 \rho ^6\right) \varepsilon _0 L^8}{126 m^4 \left(m^2+\rho ^2\right)^4 \tau ^{4/3}} \\ 
&\qquad\qquad\qquad	+ \tfrac{\left(19 m^6-35 \rho ^2 m^4-35 \rho ^4 m^2-5 \rho ^6\right) \varepsilon _0^{3/4} \eta _0 L^8}{14 m^4 \left(m^2+\rho ^2\right)^4 \tau ^2}
	+ \tfrac{\rho ^2-m^2}{\left(m^2+\rho ^2\right)^2}\biggr],\\ 
  \omega _1&=\tfrac{2 \sqrt{6} m}{L^2}+\left(-\tfrac{41 L^6 \tau^{-4/3} \varepsilon _0}{21 \sqrt{6} m^3}\right)\times \left(1-\tfrac{90 \eta _0}{41 \tau ^{2/3} \sqrt[4]{\varepsilon _0}}\right),  \displaybreak[0]\\[1.5ex] 
 (\delta a_{II2})_2&=\cF_0[\omega_2]\biggl[\tfrac{\left(178 m^8-880 \rho ^2 m^6-106 \rho ^4 m^4+315 \rho ^6 m^2+63 \rho ^8\right) \varepsilon _0 L^8}{180 m^4 \left(m^2+\rho ^2\right)^5 \tau ^{4/3}} \\ 
&\qquad\qquad\qquad	- \tfrac{\left(41 m^8-269 \rho ^2 m^6+121 \rho ^6 m^2+11 \rho ^8\right) \varepsilon _0^{3/4} \eta _0 L^8}{30 m^4 \left(m^2+\rho ^2\right)^5 \tau ^2} 
	+ \tfrac{m^4-3 \rho ^2 m^2+\rho ^4}{\left(m^2+\rho ^2\right)^3}\biggr],\\ 
  \omega _2&=\tfrac{4 \sqrt{3} m}{L^2}+\left(-\tfrac{7 \sqrt{3} L^6 \tau^{-4/3} \varepsilon _0}{10 m^3}\right)\times \left(1-\tfrac{44 \eta _0}{21 \tau ^{2/3} \sqrt[4]{\varepsilon _0}}\right),  \displaybreak[0]\\[1.5ex] 
 (\delta a_{II2})_3&=\cF_0[\omega_3]\biggl[\tfrac{\left(-4666 m^{10}+48234 \rho ^2 m^8-52030 \rho ^4 m^6-29975 \rho ^6 m^4+9978 \rho ^8 m^2+1663 \rho ^{10}\right) \varepsilon _0 L^8}{4620 m^4 \left(m^2+\rho ^2\right)^6 \tau ^{4/3}} \\ 
&\qquad\qquad\qquad	+ \tfrac{\left(211 m^{10}-2784 \rho ^2 m^8+3585 \rho ^4 m^6+1805 \rho ^6 m^4-912 \rho ^8 m^2-57 \rho ^{10}\right) \varepsilon _0^{3/4} \eta _0 L^8}{154 m^4 \left(m^2+\rho ^2\right)^6 \tau ^2} \\ 
&\qquad\qquad\qquad	+ \tfrac{-m^6+6 \rho ^2 m^4-6 \rho ^4 m^2+\rho ^6}{\left(m^2+\rho ^2\right)^4}\biggr],\\ 
  \omega _3&=\tfrac{4 \sqrt{5} m}{L^2}+\left(-\tfrac{1663 L^6 \tau^{-4/3} \varepsilon _0}{462 \sqrt{5} m^3}\right)\times \left(1-\tfrac{3420 \eta _0}{1663 \tau ^{2/3} \sqrt[4]{\varepsilon _0}}\right),  \displaybreak[0]\\[1.5ex] 
 (\delta a_{II2})_4&=\cF_0[\omega_4]\biggl[\tfrac{\left(1194 m^{12}-20697 \rho ^2 m^{10}+55709 \rho ^4 m^8-10890 \rho ^6 m^6-22928 \rho ^8 m^4+2989 \rho ^{10} m^2+427 \rho ^{12}\right) \varepsilon _0 L^8}{1170 m^4 \left(m^2+\rho ^2\right)^7 \tau ^{4/3}} \\ 
&\qquad\qquad\qquad	- \tfrac{\left(107 m^{12}-2326 \rho ^2 m^{10}+7057 \rho ^4 m^8-1840 \rho ^6 m^6-3161 \rho ^8 m^4+638 \rho ^{10} m^2+29 \rho ^{12}\right) \varepsilon _0^{3/4} \eta _0 L^8}{78 m^4 \left(m^2+\rho ^2\right)^7 \tau ^2} \\ 
&\qquad\qquad\qquad	+ \tfrac{m^8-10 \rho ^2 m^6+20 \rho ^4 m^4-10 \rho ^6 m^2+\rho ^8}{\left(m^2+\rho ^2\right)^5}\biggr],\\ 
  \omega _4&=\tfrac{2 \sqrt{30} m}{L^2}+\left(-\tfrac{427 L^6 \tau^{-4/3} \varepsilon _0}{39 \sqrt{30} m^3}\right)\times \left(1-\tfrac{870 \eta _0}{427 \tau ^{2/3} \sqrt[4]{\varepsilon _0}}\right)
\end{align*}
\subsection*{Vector meson (Type \texorpdfstring{II${}_{y}$}{II(y)})}
Vector mesons of type II${}_y$ obey a different equation of motion \eqref{minkowskiY}, such that the \acro{WKB} approximation yields a different result
\begin{align}
  A^{4d}_y(\tau) &\approx \sqrt{\frac{ \tint \omega_{4d}(\tau)\, d\tau}{\omega_{4d}\,\tau}} \, 
       \tau \, \mathcal{F}_1 .
\end{align}
The corresponding holographic ansatz 
\begin{align}
  \delta A_y &= \delta a_{IIy} \, \tau \BesselJ1( \tint \omega\, d\tau ) 
                + \delta \tilde a_{IIy} \, \tau \BesselY1( \tint \omega\, d\tau ) 
\end{align}
gives rise to the following solutions for the five lowest mesons:
\begin{align*}
(\delta a_{IIy})_0&=\tau  \cF_1[\omega_0]\biggl[\tfrac{\left(22 m^4+21 \rho ^2 m^2+7 \rho ^4\right) \varepsilon _0 L^8}{30 m^4 \left(m^2+\rho ^2\right)^3 \tau ^{4/3}} 
	- \tfrac{\left(11 m^4+4 \rho ^2 m^2+\rho ^4\right) \varepsilon _0^{3/4} \eta _0 L^8}{10 m^4 \left(m^2+\rho ^2\right)^3 \tau ^2} 
	+ \tfrac{1}{m^2+\rho ^2}\biggr],\\ 
  \omega _0&=\tfrac{2 \sqrt{2} m}{L^2}+\left(-\tfrac{7 L^6 \tau^{-4/3} \varepsilon _0}{15 \sqrt{2} m^3}\right)\times \left(1-\tfrac{6 \eta _0}{7 \tau ^{2/3} \sqrt[4]{\varepsilon _0}}\right),  \displaybreak[0]\\[1.5ex] 
 (\delta a_{IIy})_1&=\tau  \cF_1[\omega_1]\biggl[\tfrac{\left(-118 m^6+137 \rho ^2 m^4+164 \rho ^4 m^2+41 \rho ^6\right) \varepsilon _0 L^8}{126 m^4 \left(m^2+\rho ^2\right)^4 \tau ^{4/3}} \\ 
&\qquad\qquad\qquad	+ \tfrac{\left(81 m^6-105 \rho ^2 m^4-77 \rho ^4 m^2-11 \rho ^6\right) \varepsilon _0^{3/4} \eta _0 L^8}{42 m^4 \left(m^2+\rho ^2\right)^4 \tau ^2} 	+ \tfrac{\rho ^2-m^2}{\left(m^2+\rho ^2\right)^2}\biggr],\\ 
  \omega _1&=\tfrac{2 \sqrt{6} m}{L^2}+\left(-\tfrac{41 L^6 \tau^{-4/3} \varepsilon _0}{21 \sqrt{6} m^3}\right)\times \left(1-\tfrac{66 \eta _0}{41 \tau ^{2/3} \sqrt[4]{\varepsilon _0}}\right),  \displaybreak[0]\\[1.5ex] 
 (\delta a_{IIy})_2&=\tau  \cF_1[\omega_2]\biggl[\tfrac{\left(178 m^8-880 \rho ^2 m^6-106 \rho ^4 m^4+315 \rho ^6 m^2+63 \rho ^8\right) \varepsilon _0 L^8}{180 m^4 \left(m^2+\rho ^2\right)^5 \tau ^{4/3}} \\ 
&\qquad\qquad\qquad	- \tfrac{\left(129 m^8-621 \rho ^2 m^6+40 \rho ^4 m^4+209 \rho ^6 m^2+19 \rho ^8\right) \varepsilon _0^{3/4} \eta _0 L^8}{60 m^4 \left(m^2+\rho ^2\right)^5 \tau ^2} 
	+ \tfrac{m^4-3 \rho ^2 m^2+\rho ^4}{\left(m^2+\rho ^2\right)^3}\biggr],\\ 
  \omega _2&=\tfrac{4 \sqrt{3} m}{L^2}+\left(-\tfrac{7 \sqrt{3} L^6 \tau^{-4/3} \varepsilon _0}{10 m^3}\right)\times \left(1-\tfrac{38 \eta _0}{21 \tau ^{2/3} \sqrt[4]{\varepsilon _0}}\right),  \displaybreak[0]\\[1.5ex] 
 (\delta a_{IIy})_3&=\tau  \cF_1[\omega_3]\biggl[\tfrac{\left(-4666 m^{10}+48234 \rho ^2 m^8-52030 \rho ^4 m^6-29975 \rho ^6 m^4+9978 \rho ^8 m^2+1663 \rho ^{10}\right) \varepsilon _0 L^8}{4620 m^4 \left(m^2+\rho ^2\right)^6 \tau ^{4/3}} \\ 
&\qquad\qquad\qquad	+ \tfrac{\left(3449 m^{10}-34136 \rho ^2 m^8+40675 \rho ^4 m^6+15535 \rho ^6 m^4-8368 \rho ^8 m^2-523 \rho ^{10}\right) \varepsilon _0^{3/4} \eta _0 L^8}{1540 m^4 \left(m^2+\rho ^2\right)^6 \tau ^2} \\ 
&\qquad\qquad\qquad	+ \tfrac{-m^6+6 \rho ^2 m^4-6 \rho ^4 m^2+\rho ^6}{\left(m^2+\rho ^2\right)^4}\biggr],\\ 
  \omega _3&=\tfrac{4 \sqrt{5} m}{L^2}+\left(-\tfrac{1663 L^6 \tau^{-4/3} \varepsilon _0}{462 \sqrt{5} m^3}\right)\times \left(1-\tfrac{3138 \eta _0}{1663 \tau ^{2/3} \sqrt[4]{\varepsilon _0}}\right),  \displaybreak[0]\\[1.5ex] 
 (\delta a_{IIy})_4&=\tau  \cF_1[\omega_4]\biggl[\tfrac{\left(1194 m^{12}-20697 \rho ^2 m^{10}+55709 \rho ^4 m^8-10890 \rho ^6 m^6-22928 \rho ^8 m^4+2989 \rho ^{10} m^2+427 \rho ^{12}\right) \varepsilon _0 L^8}{1170 m^4 \left(m^2+\rho ^2\right)^7 \tau ^{4/3}} \\ 
&\qquad\qquad\qquad	- \tfrac{\left(891 m^{12}-14718 \rho ^2 m^{10}+40421 \rho ^4 m^8-11920 \rho ^6 m^6-14673 \rho ^8 m^4+3014 \rho ^{10} m^2+137 \rho ^{12}\right) \varepsilon _0^{3/4} \eta _0 L^8}{390 m^4 \left(m^2+\rho ^2\right)^7 \tau ^2} \\ 
&\qquad\qquad\qquad	+ \tfrac{m^8-10 \rho ^2 m^6+20 \rho ^4 m^4-10 \rho ^6 m^2+\rho ^8}{\left(m^2+\rho ^2\right)^5}\biggr],\\ 
  \omega _4&=\tfrac{2 \sqrt{30} m}{L^2}+\left(-\tfrac{427 L^6 \tau^{-4/3} \varepsilon _0}{39 \sqrt{30} m^3}\right)\times \left(1-\tfrac{822 \eta _0}{427 \tau ^{2/3} \sqrt[4]{\varepsilon _0}}\right)
\end{align*}
\defaultcolumngrid


\begin{thebibliography}{99}

\bibitem{review}
  E.~V.~Shuryak,
  ``What RHIC experiments and theory tell us about properties of  quark-gluon
  plasma?,''
  Nucl.\ Phys.\ A {\bf 750}, 64 (2005)
  [\eprint{arXiv:hep-ph/0405066}].

\bibitem{adscft} J.~M.~Maldacena,
  ``The large $N$ limit of superconformal field theories and supergravity,''
  Adv.\ Theor.\ Math.\ Phys.\  {\bf 2}, 231 (1998)
  [Int.\ J.\ Theor.\ Phys.\  {\bf 38}, 1113 (1999)]
  [\eprint{arXiv:hep-th/9711200}];\\
  S.~S.~Gubser, I.~R.~Klebanov and A.~M.~Polyakov,
  ``Gauge theory correlators from non-critical string theory,''
  Phys.\ Lett.\ B {\bf 428}, 105 (1998)
  [\eprint{arXiv:hep-th/9802109}];\\
 E.~Witten,
  ``Anti-de Sitter space and holography,''
  Adv.\ Theor.\ Math.\ Phys.\  {\bf 2}, 253 (1998)
  [\eprint{arXiv:hep-th/9802150}].


\bibitem{son}
  G.~Policastro, D.~T.~Son and A.~O.~Starinets,
  ``The shear viscosity of strongly coupled $N=4$ supersymmetric Yang-Mills
  plasma,''
  Phys.\ Rev.\ Lett.\  {\bf 87}, 081601 (2001)
  [\eprint{arXiv:hep-th/0104066}].

\bibitem{other}
  G.~Policastro, D.~T.~Son and A.~O.~Starinets,
  ``From AdS/CFT correspondence to hydrodynamics,''
  JHEP {\bf 0209}, 043 (2002)
  [\eprint{arXiv:hep-th/0205052}];\\
  P.~Kovtun, D.~T.~Son and A.~O.~Starinets,
  ``Holography and hydrodynamics: Diffusion on stretched horizons,''
  JHEP {\bf 0310}, 064 (2003)
  [\eprint{arXiv:hep-th/0309213}];\\
  P.~Kovtun, D.~T.~Son and A.~O.~Starinets,
  ``Viscosity in strongly interacting quantum field theories from black hole
  physics,''
  Phys.\ Rev.\ Lett.\  {\bf 94}, 111601 (2005)
  [\eprint{arXiv:hep-th/0405231}];\\
  A.~Buchel and J.~T.~Liu,
  ``Universality of the shear viscosity in supergravity,''
  Phys.\ Rev.\ Lett.\  {\bf 93}, 090602 (2004)
  [\eprint{arXiv:hep-th/0311175}].




\bibitem{Gubser:2006bz}
  S.~S.~Gubser,
  ``Drag force in AdS/CFT,''
  Phys.\ Rev.\  D {\bf 74}, 126005 (2006)
  [\eprint{arXiv:hep-th/0605182}].

\bibitem{Herzog:2006gh}
  C.~P.~Herzog, A.~Karch, P.~Kovtun, C.~Kozcaz and L.~G.~Yaffe,
  ``Energy loss of a heavy quark moving through $N=4$ supersymmetric
  Yang-Mills plasma,''
  JHEP {\bf 0607}, 013 (2006)
  [\eprint{arXiv:hep-th/0605158}].

\bibitem{Liu:2006ug}
  H.~Liu, K.~Rajagopal and U.~A.~Wiedemann,
  ``Calculating the jet quenching parameter from AdS/CFT,''
  Phys.\ Rev.\ Lett.\  {\bf 97}, 182301 (2006)
  [\eprint{arXiv:hep-ph/0605178}].

\bibitem{Bertoldi:2007sf}
  G.~Bertoldi, F.~Bigazzi, A.~L.~Cotrone and J.~D.~Edelstein,
  ``Holography and Unquenched Quark-Gluon Plasmas,''
  Phys.\ Rev.\  D {\bf 76}, 065007 (2007)
  [\eprint{arXiv:hep-th/0702225}].

\bibitem{Cotrone:2007qa}
  A.~L.~Cotrone, J.~M.~Pons and P.~Talavera,
  ``Notes on a SQCD-like plasma dual and holographic renormalization,''
  \eprint{arXiv:0706.2766} [hep-th].



\bibitem{KMMW}
  M.~Kruczenski, D.~Mateos, R.~C.~Myers and D.~J.~Winters,
  ``Meson spectroscopy in AdS/CFT with flavour,''
  JHEP {\bf 0307}, 049 (2003)
  [\eprint{arXiv:hep-th/0304032}].

\bibitem{ErdmengerEvans}
  J.~Babington, J.~Erdmenger, N.~J.~Evans, Z.~Guralnik and I.~Kirsch,
  ``Chiral symmetry breaking and pions in non-supersymmetric gauge /  gravity
  duals,''
  Phys.\ Rev.\  D {\bf 69}, 066007 (2004)
  [\eprint{arXiv:hep-th/0306018}].


\bibitem{Hoyos:2006gb}
  C.~Hoyos, K.~Landsteiner and S.~Montero,
  JHEP {\bf 0704}, 031 (2007)
  [arXiv:hep-th/0612169].


\bibitem{Mateos:2007vn}
  D.~Mateos, R.~C.~Myers and R.~M.~Thomson,
  ``Thermodynamics of the brane,''
  JHEP {\bf 0705}, 067 (2007)
  [\eprint{arXiv:hep-th/0701132}].


\bibitem{nastase}
  H.~Nastase,
  ``The RHIC fireball as a dual black hole,''
  \eprint{arXiv:hep-th/0501068}.

\bibitem{zahed}
  E.~Shuryak, S.~J.~Sin and I.~Zahed,
  ``A gravity dual of RHIC collisions,''
  \eprint{arXiv:hep-th/0511199}.



\bibitem{JP1}
  R.~A.~Janik and R.~Peschanski,
  ``Asymptotic perfect fluid dynamics as a consequence of AdS/CFT,''
  Phys.\ Rev.\ D {\bf 73} (2006) 045013
  [\eprint{arXiv:hep-th/0512162}].

\bibitem{RJ} R.~A.~Janik,
  ``Viscous plasma evolution from gravity using AdS/CFT,''
  Phys.\ Rev.\ Lett.\  {\bf 98}, 022302 (2007)
  [\eprint{arXiv:hep-th/0610144}].


\bibitem{SJSIN}
  S.~Nakamura and S.~J.~Sin,
  ``A holographic dual of hydrodynamics,''
  JHEP {\bf 0609}, 020 (2006)
  [\eprint{arXiv:hep-th/0607123}].

\bibitem{Heller:2007qt}
  M.~P.~Heller and R.~A.~Janik,
  ``Viscous hydrodynamics relaxation time from AdS/CFT,''
  Phys.\ Rev.\  D {\bf 76}, 025027 (2007)
  [\eprint{arXiv:hep-th/0703243}].

\bibitem{Benincasa:2007tp}
  P.~Benincasa, A.~Buchel, M.~P.~Heller and R.~A.~Janik,
  ``On the supergravity description of boost invariant conformal plasma at
  strong coupling,''
  arXiv:0712.2025 [hep-th].

\bibitem{JP2}
  R.~A.~Janik and R.~Peschanski,
   ``Gauge / gravity duality and thermalization of a boost-invariant perfect
  fluid,''
  Phys.\ Rev.\ D {\bf 74} (2006) 046007
  [\eprint{arXiv:hep-th/0606149}].

\bibitem{BAK}
  D.~Bak and R.~A.~Janik,
  ``From static to evolving geometries: R-charged hydrodynamics from
  supergravity,''
  Phys.\ Lett.\  B {\bf 645} (2007) 303
  [\eprint{arXiv:hep-th/0611304}].

\bibitem{SJSIN2}
  S.~J.~Sin, S.~Nakamura and S.~P.~Kim,
  ``Elliptic flow, Kasner universe and holographic dual of RHIC fireball,''
  JHEP {\bf 0612}, 075 (2006)
  [\eprint{arXiv:hep-th/0610113}].



\bibitem{Kajantie}
  K.~Kajantie and T.~Tahkokallio,
  ``Spherically expanding matter in AdS/CFT,''
  \eprint{arXiv:hep-th/0612226}.

\bibitem{Siopsis}
  J.~Alsup, C.~Middleton and G.~Siopsis,
  ``AdS/CFT Correspondence with Heat Conduction,''
  \eprint{arXiv:hep-th/0607139}.

\bibitem{Kovchegov}
  Y.~V.~Kovchegov and A.~Taliotis,
  ``Early time dynamics in heavy ion collisions from AdS/CFT correspondence,''
  Phys.\ Rev.\  C {\bf 76}, 014905 (2007)
  [\eprint{arXiv:0705.1234 [hep-ph}].

\bibitem{Bjorken}
 J.~D.~Bjorken,
  ``Highly Relativistic Nucleus-Nucleus Collisions: The Central Rapidity
  Region,''
  Phys.\ Rev.\ D {\bf 27}, 140 (1983).

\bibitem{hydro}
  P.~Huovinen and P.~V.~Ruuskanen,
  ``Hydrodynamic models for heavy ion collisions,''
  \eprint{arXiv:nucl-th/0605008};\\
 P.~F.~Kolb and U.~W.~Heinz,
  ``Hydrodynamic description of ultrarelativistic heavy-ion collisions,''
  \eprint{arXiv:nucl-th/0305084}.


\bibitem{KarchOBannon}
  A.~Karch and A.~O'Bannon,
  ``Chiral transition of $N = 4$ super Yang-Mills with flavor on a 3-sphere,''
  Phys.\ Rev.\  D {\bf 74} (2006) 085033
  [\eprint{arXiv:hep-th/0605120}].


\bibitem{KarchKatz}
  A.~Karch and E.~Katz,
  ``Adding flavor to AdS/CFT,''
  JHEP {\bf 0206}, 043 (2002)
  [\eprint{arXiv:hep-th/0205236}].

\bibitem{Karch:2005ms}
  A.~Karch, A.~O'Bannon and K.~Skenderis,
  ``Holographic renormalization of probe D-branes in AdS/CFT,''
  JHEP {\bf 0604} (2006) 015
  [\eprint{arXiv:hep-th/0512125}].

\bibitem{Skenderis}
S.~de Haro, S.~N.~Solodukhin and K.~Skenderis,
  ``Holographic reconstruction of spacetime and renormalization in the
  AdS/CFT correspondence,''
  Commun.\ Math.\ Phys.\  {\bf 217}, 595 (2001)
  [\eprint{arXiv:hep-th/0002230}];\\
 K.~Skenderis,
  ``Lecture notes on holographic renormalization,''
  Class.\ Quant.\ Grav.\  {\bf 19}, 5849 (2002)
  [\eprint{arXiv:hep-th/0209067}].

\bibitem{Kasper}
  K.~Peeters, J.~Sonnenschein and M.~Zamaklar,
  Phys.\ Rev.\  D {\bf 74}, 106008 (2006)
  [arXiv:hep-th/0606195].

\end{thebibliography}
\end{document}